\documentclass[apj, twocolappendix]{emulateapj}

\usepackage{graphicx}
\usepackage{amsmath}
\usepackage{natbib}

\usepackage{color}

\begin{document}

\title{
  The gravitational interaction between planets on inclined orbits and
  protoplanetary disks as the origin of primordial spin--orbit misalignments
}

\author{Titos Matsakos and Arieh K\"onigl}

\affil{
  Department of Astronomy \& Astrophysics and The Enrico Fermi Institute,
  The University of Chicago, Chicago, IL 60637, USA
}

\shortauthors{Matsakos \& K\"onigl}
\shorttitle{Planet-induced spin--orbit misalignments}

\begin{abstract}
Many of the observed spin--orbit alignment properties of exoplanets can be
explained in the context of the primordial disk misalignment model, in which an
initially aligned protoplanetary disk is torqued by a distant stellar companion
on a misaligned orbit, resulting in a precessional motion that can lead to
large-amplitude oscillations of the spin--orbit angle.
We consider a variant of this model in which the companion is a giant planet
with an orbital radius of a few au.
Guided by the results of published numerical simulations, we model the dynamical
evolution of this system by dividing the disk into inner and outer
parts---separated at the location of the planet---that behave as distinct, rigid
disks.
We show that the planet misaligns the inner disk even as the orientation of the
outer disk remains unchanged.
In addition to the oscillations induced by the precessional motion, whose
amplitude is larger the smaller the initial inner-disk-to-planet mass ratio, the
spin--orbit angle also exhibits a secular growth in this case---driven by
ongoing mass depletion from the disk---that becomes significant when the inner
disk's angular momentum drops below that of the planet.
Altogether, these two effects can produce significant misalignment angles for
the inner disk, including retrograde configurations.
We discuss these results within the framework of the Stranded Hot Jupiter
scenario and consider their implications, including to the interpretation of the
alignment properties of debris disks.
\end{abstract}


\maketitle

\section{Introduction}
  \label{sec:intro}

A major open question in the study of exoplanets is the origin of their apparent
obliquity properties---the distribution of the angle $\lambda$ between the
stellar spin and the planet's orbital angular momentum vectors as projected on
the sky (see, e.g., the review by \citealt{WinnFabrycky15}).
Measurements of the Rossiter--McLaughlin effect in hot Jupiters (HJs, defined
here as planets with masses $M_\mathrm{p}\gtrsim0.3\,M_\mathrm{J}$ that have
orbital periods $P_\mathrm{orb} \lesssim 10\,$days) have indicated that
$\lambda$ spans the entire range from~$0^\circ$ to~$180^\circ$, in stark
contrast with the situation in the solar system (where the angle between the
planets' total angular momentum vector and that of the Sun is only
$\sim$$6^\circ$).
In addition, there is a marked difference in the distribution of $\lambda$
between G~stars, where $\sim$$1/2$ of systems are well aligned
($\lambda < 20^\circ$) and the rest are spread out roughly uniformly over the
remainder of the $\lambda$ range, and F~stars of effective temperature
$T_\mathrm{eff} \gtrsim 6250\,$K, which exhibit only a weak excess of
well-aligned systems. There is, however, also evidence for a dependence of the 
obliquity distribution on the properties of the planets and not just on those of
the host star; in particular, only planets with $M_\mathrm{p} < 3\,M_\mathrm{J}$ 
have apparent retrograde orbits ($\lambda > 90^\circ$). 

Various explanations have been proposed to account for the broad range of
observed obliquities, but the inferred dependences on $T_\mathrm{eff}$ and
$M_\mathrm{p}$ provide strong constraints on a viable model. In one scenario 
\cite[][]{Winn+10, Albrecht+12}, HJs arrive in the vicinity of the host star on
a misaligned orbit and subsequently act to realign the host through a tidal 
interaction, which is more effective in cool stars than in hot ones.
In this picture, HJs form at large radii and either migrate inward through their
natal disk while maintaining nearly circular orbits or are placed on a
high-eccentricity orbit after the gaseous disk dissipates---which enables them
to approach the center and become tidally trapped by the star (with their orbits
getting circularized by tidal friction; e.g., \citealt{FordRasio06}).\footnote{
The possibility of HJs forming at their observed locations has also been
considered in the literature \citep[e.g.,][]{Boley+16,Batygin+16}, but the
likelihood of this scenario is still being debated.}
The processes that initiate high-eccentricity migration (HEM), which can be
either planet--planet scattering \citep[e.g.,][]{Chatterjee+08, JuricTremaine08,
BeaugeNesvorny12} or secular interactions that involve a stellar binary
companion or one or more planetary companions (such as Kozai-Lidov oscillations
--- e.g., \citealt{WuMurray03, FabryckyTremaine07, Naoz+11, Petrovich15b}---and
secular chaos---e.g., \citealt{WuLithwick11, LithwickWu14, Petrovich15a,
Hamers+17}), all give rise to HJs with a distribution of misaligned orbits.
In the case of classical disk migration, the observed obliquities can be
attributed to a primordial misalignment of the natal disk that occurred during
its initial assembly from a turbulent interstellar gas \citep[e.g.,][]{Bate+10,
Fielding+15} or as a result of magnetic and/or gravitational torques induced,
respectively, by a tilted stellar dipolar field and a misaligned companion
\citep[e.g.,][]{Lai+11, Batygin12, BatyginAdams13, Lai14, SpaldingBatygin14}.

The tidal realignment hypothesis that underlies  the above modeling framework
was challenged by the results of \citet{Mazeh+15}, who examined the rotational
photometric modulations of a large number of {\it Kepler}\/ sources.
Their analysis indicated that the common occurrence of aligned systems around
cool stars characterizes the general population of planets and not just HJs,
and, moreover, that this property extends to orbital periods as long as
$\sim$$50\,$days, about an order of magnitude larger than the maximum value of
$P_\mathrm{orb}$ for which tidal interaction with the star remains important.
To reconcile this finding with the above scenario, \citet{MatsakosKonigl15}
appealed to the results of planet formation and evolution models, which predict
that giant planets form efficiently in protoplanetary disks and that most of
them migrate rapidly to the disk's inner edge, where, if the arriving planet's
mass is not too high ($\lesssim 1\,M_\mathrm{J}$), it could remain stranded near
that radius for up to $\sim$$1\,$Gyr---until it gets tidally ingested by the
host star.
They proposed that the ingestion of a stranded HJ (SHJ)---which is accompanied
by the transfer of its orbital angular momentum to the star---is the dominant
spin-realignment mechanism.
In this picture, the dichotomy in the obliquity properties between cool and hot
stars is a direct consequence of the higher efficiency of magnetic braking and
lower moment of inertia of the former in comparison with the latter.
By applying a simple dynamical model to the observed HJ distributions in~G and
F~stars, \citet{MatsakosKonigl15} inferred that $\sim$50\% of planetary systems
harbor an SHJ with a typical mass of $\sim$$0.6\,M_\mathrm{J}$.
In this picture, the obliquity properties of currently observed HJs---and the
fact that they are consistent with those of lower-mass and more distant
planets---are most naturally explained if most of the planets in a given
system---including any SHJ that  may have been present---are formed in, and
migrate along the plane of, a primordially misaligned disk.\footnote{
This explanation does not necessarily imply that all planets that reached the
vicinity of the host star must have moved in by classical migration, although
SHJs evidently arrived in this way.
In fact, \citet{MatsakosKonigl16} inferred that most of the planets that
delineate the boundary of the so-called sub-Jovian desert in the
orbital-period--planet-mass plane got in by a secular HEM process (one that,
however, did not give rise to high orbital inclinations relative to the natal
disk plane).}
This interpretation is compatible with the properties of systems like Kepler-56,
in which two close-in planets have $\lambda \approx 45^\circ$ and yet are nearly
coplanar \citep{Huber+13}, and 55~Cnc, a coplanar five-planet system with
$\lambda \approx 72^\circ$ \citep[e.g.,][]{Kaib+11, BourrierHebrard14}.\footnote{
The two-planet system KOI-89 \citep{Ahlers+15} may be yet another example.}
It is also consistent with the apparent lack of a correlation between the
obliquity properties of observed HJs and the presence of a massive companion
\citep[e.g.,][]{Knutson+14, Ngo+15, Piskorz+15}.

In this paper we explore a variant of the primordial disk misalignment model
first proposed by \citet{Batygin12}, in which, instead of the tilting of the
entire disk by a distant ($\sim$500\,au) stellar companion on an inclined orbit,
we consider the gravitational torque exerted by a much closer ($\sim$5\,au)
\emph{planetary} companion on such an orbit, which acts to misalign \emph{only
the inner region} of the protoplanetary disk.
This model is motivated by the inferences from radial velocity surveys and
adaptive-optics imaging data (\citealt{Bryan+16}; see also \citealt{Knutson+14}) 
that $\sim$70\% of planetary systems harboring a transiting HJ have a companion 
with mass in the range 1--13\,$M_\mathrm{J}$ and semimajor axis in the range 
$1$--$20$\,au, and that $\sim$50\% of systems harboring one or two planets 
detected by the radial velocity method have a companion with mass in the range 
$1$--$20\,M_\mathrm{J}$ and semimajor axis in the range $5$--$20$\,au.
Further motivation is provided by the work of \citet{LiWinn16}, who re-examined
the photometric data analyzed by \citet{Mazeh+15} and found indications that the
good-alignment property of planets around cool stars does not hold for large
orbital periods, with the obliquities of planets with
$P_\mathrm{orb} \gtrsim 10^2\,$days appearing to tend toward a random
distribution.

One possible origin for a giant planet on an inclined orbit with a semimajor
axis $a$ of a few au is planet--planet scattering in the natal disk.
Current theories suggest that giant planets may form in tightly packed
configurations that can become dynamically unstable and undergo orbit crossing
(see, e.g., \citealt{Davies+14} for a review).
The instabilities start to develop before the gaseous disk component dissipates
\citep[e.g.,][]{Matsumura+10, Marzari+10}, and it has been argued
\citep{Chatterjee+08} that the planet--planet scattering process may, in fact,
peak before the disk is fully depleted of gas (see also \citealt{Lega+13}).
A close encounter between two giant planets is likely to result in a collision
if the ratio $(M_\mathrm{p}/M_*)(a/R_\mathrm{p})$ (the Safronov number) is $< 1$
(where $M_*$ is the stellar mass and $R_\mathrm{p}$ is the planet's radius), and
in a scattering if this ratio is $> 1$ \citep[e.g.,][]{FordRasio08}.
The scattering efficiency is thus maximized when a giant planet on a
comparatively wide orbit is involved \citep[cf.][]{Petrovich+14}.
High inclinations might also be induced by resonant excitation in giant planets
that become trapped in a mean-motion resonance through classical (Type II) disk
migration \citep{ThommesLissauer03, LibertTsiganis09}, and this process could,
moreover, provide an alternative pathway to planet--planet scattering
\citep{LibertTsiganis11}.
In these scenarios, the other giant planets that were originally present in the 
disk can be assumed to have either been ejected from the system in the course of
their interaction with the remaining misaligned planet or else reached the star
at some later time through disk migration.
As we show in this paper, a planet on an inclined orbit can have a significant
effect on the orientation of the disk region interior to its orbital radius when
the mass of that region decreases to the point where the inner disk's angular
momentum becomes comparable to that of the planet.
For typical mass depletion rates in protoplanetary disks \citep[e.g.,][]
{BatyginAdams13}, this can be expected to happen when the system's age is
$\sim$$10^6$--$10^7\,$yr, which is comparable to the estimated formation time of
Jupiter-mass planets at $\gtrsim 5\,$au.
In the proposed scenario, a planet of mass $M_\mathrm{p} \gtrsim M_\mathrm{J}$
is placed on a high-inclination orbit at a time $t_0 \gtrsim 1\,$Myr that, on
the one hand, is late enough for the disk mass interior to the planet's location
to have decreased to a comparable value, but that, on the other hand, is early
enough for the inner disk to retain sufficient mass after becoming misaligned to
enforce the orbital misalignment of existing planets and/or form new planets in
its reoriented orbital plane (including any Jupiter-mass planets destined to
become an HJ or an SHJ).

The dynamical model adopted in this paper is informed by the
smooth-particle-hydrodynamics simulations carried out by
\citet{Xiang-GruessPapaloizou13}.
They considered the interaction between a massive ($1$--$6\,M_\mathrm{J}$)
planet that is placed on an inclined, circular orbit of radius $5$\,au and a
low-mass ($0.01\,M_*$) protoplanetary disk that extends to $25$\,au.
A key finding of these simulations was that the disk develops a warped
structure, with the regions interior and exterior to the planet's radial
location behaving as separate, rigid disks with distinct inclinations; in
particular, the inner disk was found to exhibit substantial misalignment with
respect to its initial direction when the planet's mass was large enough and its
initial inclination was intermediate between the limits of $0^\circ$ and
$90^\circ$ at which no torque is exerted on the disk.
Motivated by these results, we construct an analytic model for the gravitational
interaction between the planet and the two separate parts of the disk.
The general effect of an interaction of this type between a planet on an
inclined orbit and a rigid disk is to induce a precession of the planet's orbit
about the total angular momentum vector.
In contrast with \citet{Xiang-GruessPapaloizou13}, whose simulations only
extended over a fraction of a precession period, we consider the long-term
evolution of such systems.
In particular, we use our analytic model to study how the ongoing depletion of
the disk's mass affects the orbital orientations of the planet and of the disk's
two parts.
We describe the model in Section~\ref{sec:model} and present our calculations in
Section~\ref{sec:results}.
We discuss the implications of these results to planet obliquity measurements
and to the alignment properties of debris disks in Section~\ref{sec:discussion},
and summarize in Section~\ref{sec:conclusion}.

\section{Modeling approach}
  \label{sec:model}

\subsection{Assumptions}
\label{subsec:assumptions}

\begin{figure}
  \includegraphics[width=\columnwidth]{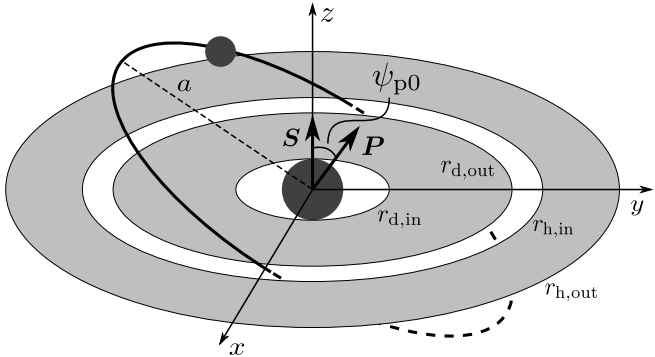}
  \caption{
    Schematic representation (not to scale) of the initial configuration of our
    model.
    See text for details.
    \label{fig:initial}}
\end{figure}

The initial configuration that we adopt is sketched in Figure~\ref{fig:initial}. 
We consider a young star (subscript s) that is surrounded by a Keplerian
accretion disk, and a Jupiter-mass planet (subscript p) on a circular orbit.
The disk consists of two parts: an inner disk (subscript d) that extends between
an inner radius $r_\mathrm{d,in}$ and an outer radius $r_\mathrm{d,out}$, and an
outer disk (subscript h) that extends between $r_\mathrm{h,in}$ and
$r_\mathrm{h,out}$; they are separated by a narrow gap that is centered on the
planet's orbital radius $a$.
The two parts of the disk are initially coplanar, with their normals aligned
with the stellar angular momentum vector $\boldsymbol{S}$, whereas the planet's
orbital angular momentum vector $\boldsymbol{P}$ is initially inclined at an
angle $\psi_\mathrm{p0}$ with respect to $\boldsymbol{S}$ (where the subscript
$0$ denotes the time $t = t_0$ at which the planet is placed on the inclined
orbit).

We assume that, during the subsequent evolution, each part of the disk maintains
a flat geometry and precesses as a rigid body.
The rigidity approximation is commonly adopted in this context and is attributed
to efficient communication across the disk through the propagation of bending
waves or the action of a viscous stress (e.g., \citealt{Larwood+96}; see also
\citealt{Lai14} and references therein).\footnote{
One should, however, bear in mind that real accretion disks are inherently fluid
in nature and therefore cannot strictly obey the rigid-body approximation; see,
e.g., \citet{Rawiraswattana+16}.}
Based on the simulation results presented in \citet{Xiang-GruessPapaloizou13},
we conjecture that this communication is severed at the location of the planet.
This outcome is evidently the result of the planet's opening up a gap in the
disk, although it appears that the gap need not be fully evacuated for this
process to be effective.
In fact, the most strongly warped simulated disk configurations correspond to
comparatively high initial inclination angles, for which the planet spends a
relatively small fraction of the orbital time inside the disk, resulting in gaps
that are less deep and wide than in the fully embedded case.
Our calculations indicate that, during the disk's subsequent evolution, its
inner and outer parts may actually detach as a result of the precessional
oscillation of the inner disk.
This oscillation is particularly strong in the case of highly mass-depleted
disks on which we focus attention in this paper: in the example shown in
Figure~\ref{fig:all-m} below, the initial amplitude of this oscillation is
$\sim$$40^\circ$.

The planet's orbital inclination is subject to damping by dynamical friction
\citep{Xiang-GruessPapaloizou13}, although the damping rate is likely low for
the high values of $\psi_\mathrm{p0}$ that are of particular interest to us
\citep{Bitsch+13}.
Furthermore, in cases where the precessional oscillation of the inner disk
causes the disk to split at the orbital radius of the planet, one can plausibly
expect the local gas density to become too low for dynamical friction to
continue to play a significant role on timescales longer than the initial
oscillation period ($\sim$$10^4$\,yr for the example shown in
Figure~\ref{fig:all-m}).
In light of these considerations, and in the interest of simplicity, we do not
include the effects of dynamical friction in any of our presented models.

As a further simplification, we assume that the planet's orbit remains circular.
The initial orbital eccentricity of a planet ejected from the disk by either of
the two mechanisms mentioned in Section~\ref{sec:intro} may well have a
nonnegligible eccentricity.
However, the simulations performed by \citet{Bitsch+13} indicate that the
dynamical friction process damps eccentricities much faster than inclinations,
so that the orbit can potentially be circularized on a timescale that is shorter
than the precession time (i.e., before the two parts of the disk can become
fully separated).
On the other hand, even if the initial eccentricity is zero, it may be pumped up
by the planet's gravitational interaction with the outer disk if
$\psi_\mathrm{p0}$ is high enough ($\gtrsim 20^\circ$;
\citealt{Teyssandier+13}).
This is essentially the Kozai-Lidov effect, wherein the eccentricity undergoes
periodic oscillations in antiphase with the orbital inclination 
\citep{TerquemAjmia10}.
These oscillations were noticed in the numerical simulations of
\citet{Xiang-GruessPapaloizou13} and \citet{Bitsch+13}.
Their period can be approximated by $\tau_\mathrm{KL} \sim (r_\mathrm{h,out}/
r_\mathrm{h,in})^2 (2\pi/|\Omega_\mathrm{ph}|)$ \citep{TerquemAjmia10}, where we
used the expression for the precession frequency $\Omega_\mathrm{ph}$
(Equation~(\ref{eq:omega_ph})) that corresponds to the torque exerted by the
outer disk on the misaligned planet.
For the parameters of the representative mass-depleted disk model shown in
Figure~\ref{fig:all-m}, $\tau_\mathrm{KL} \sim 10^6$\,yr.
This time is longer by a factor of $\sim$$10^2$ than the initial precession
period of the inner disk in this example, implying that the Kozai-Lidov process
will have little effect on the high-amplitude oscillations of $\psi_\mathrm{p}$.
Kozai-Lidov oscillations might, however, modify the details of the long-term
behavior of the inner disk, since $\tau_\mathrm{KL}$ is comparable to the
mass-depletion time $\tau$ (Equation~(\ref{eq:deplete})) that underlies the
secular evolution of the system.

Our model takes into account the tidal interaction of the spinning star with the
inner and outer disks and with the planet, which was not considered in the
aforementioned simulations.
The inclusion of this interaction is motivated by the finding
\citep{BatyginAdams13, Lai14, SpaldingBatygin14} that an evolving protoplanetary
disk with a binary companion on an inclined orbit can experience a resonance
between the disk precession frequency (driven by the companion) and the stellar
precession frequency (driven by the disk), and that this resonance crossing can
generate a strong misalignment between the angular momentum vectors of the disk
and the star.
As it turns out (see Section~\ref{sec:results}), in the case that we
consider---in which the companion is a Jupiter-mass planet with an orbital
radius of a few au rather than a solar-mass star at a distance of a few hundred
au---this resonance is not encountered.
We also show that, even in the case of a binary companion, the misalignment
effect associated with the resonance crossing is weaker than that inferred in
the above works when one also takes into account the torque that the \emph{star}
exerts on the inner disk (see Appendix~\ref{app:resonance}).

\subsection{Equations}

We model the dynamics of the system by following the temporal evolution of the
angular momenta ($\boldsymbol{S}$, $\boldsymbol{D}$, $\boldsymbol{P}$, and
$\boldsymbol{H}$) of the four constituents (the star, the inner disk, the
planet, and the outer disk, respectively) due to their mutual gravitational
torques.
Given that the orbital period of the planet is much shorter than the
characteristic precession time scales of the system, we approximate the planet
as a ring of uniform density, with a total mass equal to that of the planet and
a radius equal to its semimajor axis.

The evolution of the angular momentum $\boldsymbol L_k$ of an object $k$ under
the influence of a torque $\boldsymbol T_{ik}$ exerted by an object $i$ is given
by $d\boldsymbol L_k/dt = \boldsymbol T_{ik}$.
The set of equations that describes the temporal evolution of the four angular
momenta is thus
\begin{equation}
  \frac{d\boldsymbol S}{dt} = \boldsymbol T_\mathrm{ds}
    + \boldsymbol T_\mathrm{ps} + \boldsymbol T_\mathrm{hs}\,,
\end{equation}
\begin{equation}
  \frac{d\boldsymbol D}{dt} = \boldsymbol T_\mathrm{sd}
    + \boldsymbol T_\mathrm{pd} + \boldsymbol T_\mathrm{hd}\,,
\end{equation}
\begin{equation}
  \frac{d\boldsymbol P}{dt} = \boldsymbol T_\mathrm{sp}
    + \boldsymbol T_\mathrm{dp} + \boldsymbol T_\mathrm{hp}\,,
\end{equation}
\begin{equation}
  \frac{d\boldsymbol H}{dt} = \boldsymbol T_\mathrm{sh}
    + \boldsymbol T_\mathrm{dh} + \boldsymbol T_\mathrm{ph}\,,
\end{equation}
where $\boldsymbol T_{ik} = -\boldsymbol T_{ki}$.
The above equations can also be expressed in terms of the precession frequencies
$\Omega_{ik}$:
\begin{equation}
  \frac{d\boldsymbol L_k}{dt}
    = \sum_i\boldsymbol T_{ik}
    = \sum_i\Omega_{ik}\frac{\boldsymbol L_i\times\boldsymbol L_k}{J_{ik}}\,,
  \label{eq:precession}
\end{equation}
where $J_{ik} = |\boldsymbol L_i + \boldsymbol L_k|
= (L_i^2 + L_k^2 + 2L_iL_k\cos{\theta_{ik}})^{1/2}$ and
$\Omega_{ik} = \Omega_{ki}$.
In Appendix~\ref{app:torques} we derive analytic expressions for the torques
$\boldsymbol T_{ik}$ and the corresponding precession frequencies $\Omega_{ik}$.

\subsection{Numerical Setup}

The host is assumed to be a protostar of mass $M_* = M_\odot$,
radius $R_* = 2R_\odot$, rotation rate $\Omega_* = 0.1(GM_*/R_*^3)^{1/2}$, and
angular momentum
\begin{eqnarray}
  S &=& k_*M_*R_*^2\Omega_* =
  1.71 \times 10^{50}\\
  &\times& \left(\frac{k_*}{0.2}\right) \left(\frac{M_*}{M_\odot}\right)
  \left(\frac{R_*}{2R_\odot}\right)^2
  \left(\frac{\Omega_*}{0.1\sqrt{GM_\odot/(2R_\odot)^3}}\right)\,
  \mathrm{erg\,s}\nonumber\,,
\end{eqnarray}
where $k_* \simeq 0.2$ for a fully convective star (modeled as a polytrope of
index $n = 1.5$).
The planet is taken to have Jupiter's mass and radius,
$M_\mathrm{p} = M_\mathrm{J}$ and $R_\mathrm{p} = R_\mathrm{J}$, and a
fixed semimajor axis, $a = 5$\,au, so that its orbital angular momentum is
\begin{eqnarray}
  P &=& M_\mathrm{p}(GM_*a)^{1/2} =
  1.89 \times 10^{50}
  \label{eq:P}\\
  &&\times \left(\frac{M_\mathrm{p}}{M_\mathrm{J}}\right)
  \left(\frac{M_*}{M_\odot}\right)^{1/2}
  \left(\frac{a}{5\,\mathrm{au}}\right)^{1/2}\,\mathrm{erg\,s}\,.\nonumber
\end{eqnarray}

We consider two values for the total initial disk mass: (1)
$M_\mathrm{t0} = 0.1\,M_*$, corresponding to a comparatively massive disk, and
(2) $M_\mathrm{t0} = 0.02\,M_*$, corresponding to a highly evolved system that
has entered the transition-disk phase.
In both cases we take the disk surface density to scale with radius as $r^{-1}$.
The inner disk extends from $r_\mathrm{d,in} = 4R_\odot$ to
$r_\mathrm{d,out} = a$, and initially has $10\%$ of the total mass.
Its angular momentum is
\begin{eqnarray}
  D &=& \frac{2}{3}M_\mathrm{d}\left(GM_*\right)^{1/2}
    \frac{r_\mathrm{d,out}^{3/2} - r_\mathrm{d,in}^{3/2}}
    {r_\mathrm{d,out} - r_\mathrm{d,in}}
  \label{eq:D}\\
  &\simeq& 1.32 \times 10^{51}\,
  \left(\frac{M_\mathrm{d}}{0.01M_\odot}\right)
  \left(\frac{M_*}{M_\odot}\right)^{1/2}
  \left(\frac{a}{5\,\mathrm{au}}\right)^{1/2}\, \mathrm{erg\,s} \nonumber \,.
\end{eqnarray}
The outer disk has edges at $r_\mathrm{h,in} = a$ and
$r_\mathrm{h,out} = 50$\,au, and angular momentum
\begin{eqnarray}
  H &=& \frac{2}{3}M_\mathrm{h}\left(GM_*\right)^{1/2}
    \frac{r_\mathrm{h,out}^{3/2} - r_\mathrm{h,in}^{3/2}}
    {r_\mathrm{h,out} - r_\mathrm{h,in}}\\
 &\simeq& 3.76 \times 10^{52}\,
 \left(\frac{M_\mathrm{h}}{0.09M_\odot}\right)
  \left(\frac{M_*}{M_\odot}\right)^{1/2}
  \left(\frac{r_\mathrm{h,out}}{50\,\mathrm{au}}\right)^{1/2}\, \mathrm{erg\,s}
  \nonumber\,.
\end{eqnarray}

We model mass depletion in the disk using the expression first employed in this
context by \citet{BatyginAdams13},
\begin{equation}
  M_\mathrm{t}(t) = \frac{M_{\mathrm{t}}(t=0)}{1 + t/\tau}\,,
  \label{eq:deplete}
\end{equation}
where we adopt $M_{\mathrm{t}}(t=0)=0.1\,M_\sun$ and $\tau = 0.5$\,Myr as in
\citet{Lai14}.
We assume that this expression can also be applied separately to the inner and
outer parts of the disk.
The time evolution of the inner disk's angular momentum due to mass depletion is
thus given by
\begin{equation}
  \label{eq:dDdt}
  \left(\frac{d\boldsymbol{D}}{dt}\right)_\mathrm{depl}
    = -\frac{D_0}{\tau(1 + t/\tau)^2}\hat{\boldsymbol{D}}
    = -\frac{\boldsymbol{D}}{\tau+t}\,.
\end{equation}
For the outer disk we assume that the presence of the planet inhibits efficient
mass accretion, and we consider the following limits: (1) the outer disk's mass
remains constant, and (2) the outer disk loses mass (e.g., through
photoevaporation) at the rate given by Equation~(\ref{eq:deplete}).\footnote{
After the inner disk tilts away from the outer disk, the inner rim of the outer
disk becomes exposed to the direct stellar radiation field, which accelerates
the evaporation process \citep{Alexander+06}.
According to current models, disk evaporation is induced primarily by X-ray and
FUV photons and occurs at a rate of
$\sim$$10^{-9}$--$10^{-8}\,M_\sun\,\mathrm{yr}^{-1}$ for typical stellar
radiation fields (see \citealt{Gorti+16} for a review).
Even if the actual rate is near the lower end of this range, the outer disk in
our low-$M_{\rm t0}$ models would be fully depleted of mass on a timescale of
$\sim$$10$\,Myr; however, a similar outcome for the high-$M_\mathrm{t0}$ models
would require the mass evaporation rate to be near the upper end of the
estimated range.}
We assume that any angular momentum lost by the disk is transported out of the
system (for example, by a disk wind).

We adopt a Cartesian coordinate system ($x,\,y,\,z$) as the ``lab'' frame of
reference (see Figure~\ref{fig:initial}).
Initially, the equatorial plane of the star and the planes of the inner and
outer disks coincide with the $x$--$y$ plane (i.e.,
$\psi_\mathrm{s0} = \psi_\mathrm{d0} = \psi_\mathrm{h0} = 0$, where $\psi_k$
denotes the angle between $\boldsymbol{L}_k$ and the $z$ axis), and only the
orbital plane of the planet has a finite initial inclination
($\psi_\mathrm{p0}$).
The $x$ axis is chosen to coincide with the initial line of nodes of the
planet's orbital plane.

\begin{table*}
\begin{center}
\caption{Model parameters\label{tab:models}}
\begin{tabular}{l|cccccccccc}
\hline\hline
Model & $\boldsymbol{S}$ & $\boldsymbol{D}$ & $\boldsymbol{P}$ & $\boldsymbol{H}$
           & $M_\mathrm{d0} \ [M_*] $ & $M_\mathrm{h0}\ [M_*]$ & $M_\mathrm{t0}\ [M_*]$
           & $M_\mathrm{p}$ & $a$ [au] & $\psi_\mathrm{p0}\ [^\circ]$ \\
\hline
\texttt{DP-M}       & --      & $\surd$ & $\surd$ & --      & $0.010\downarrow$ & --                & --                     & $M_\mathrm{J}$ & $5$   &  $60$  \\
\texttt{DP-m}       & --      & $\surd$ & $\surd$ & --      & $0.002\downarrow$ & --                & --                     & $M_\mathrm{J}$ & $5$   &  $60$  \\
\texttt{all-M}      & $\surd$ & $\surd$ & $\surd$ & $\surd$ & $0.010\downarrow$ & $0.090\downarrow$ & $0.10$                 & $M_\mathrm{J}$ & $5$   &  $60$  \\
\texttt{all-m}      & $\surd$ & $\surd$ & $\surd$ & $\surd$ & $0.002\downarrow$ & $0.018\downarrow$ & $0.02$                 & $M_\mathrm{J}$ & $5$   &  $60$  \\
\texttt{all-Mx}     & $\surd$ & $\surd$ & $\surd$ & $\surd$ & $0.010\downarrow$ & $0.090$ --        & $0.10$                 & $M_\mathrm{J}$ & $5$   &  $60$  \\
\texttt{all-mx}     & $\surd$ & $\surd$ & $\surd$ & $\surd$ & $0.002\downarrow$ & $0.018$ --        & $0.02$                 & $M_\mathrm{J}$ & $5$   &  $60$  \\
\texttt{retrograde} & $\surd$ & $\surd$ & $\surd$ & $\surd$ & $0.002\downarrow$ & $0.018\downarrow$ & $0.02$                 & $M_\mathrm{J}$ & $5$   & $110$  \\
\texttt{binary}     & $\surd$ & $\surd$ & $\surd$ & --      & --                & --                & $\ \ \,0.10\downarrow$ & $M_\odot$      & $300$ &  $10$  \\
\hline
\end{tabular}
\end{center}
\end{table*}
Table~\ref{tab:models} presents the models we explore and summarizes the
relevant parameters.
Specifically, column 1 contains the models' designations (with the letters
\texttt{M} and \texttt{m} denoting, respectively, high and low disk masses at
time $t=t_0$), columns 2--5 indicate which system components are being
considered, columns 6--9 list the disk and planet masses (with the arrow
indicating active mass depletion), and columns 10 and~11 give the planet's
semimajor axis and initial misalignment angle, respectively.
The last listed model (\texttt{binary}) does not correspond to a planet
misaligning the inner disk but rather to a binary star tilting the entire disk.
This case is considered for comparison with the corresponding model in
\citet{Lai14}.

\section{Results}
  \label{sec:results}

The gravitational interactions among the different components of the system that
we consider (star, inner disk, planet, and outer disk) can result in a highly
nonlinear behavior.
To gain insight into these interactions we start by analyzing a much simpler
system, one consisting only of the inner disk and the (initially misaligned)
planet.
The relevant timescales that characterize the evolution of this system are the
precession period $\tau_\mathrm{dp} \equiv 2\pi/\Omega_\mathrm{dp}$
(Equation~(\ref{eq:omega_dp})) and the mass depletion timescale
$\tau = 5\times 10^5\,$yr (Equation~(\ref{eq:deplete})).

\begin{figure*}
  \includegraphics[width=\textwidth]{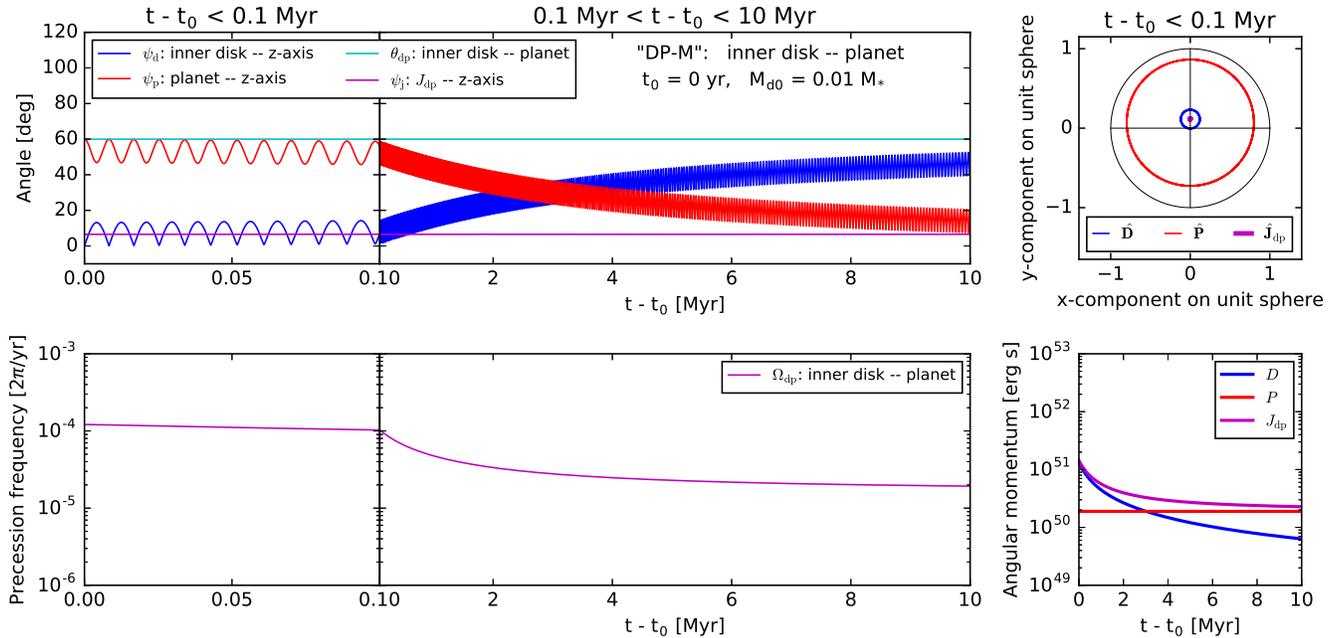}
  \caption{
    Time evolution of a ``reduced'' system, consisting of just a planet and an
    inner disk, for an initial disk mass  $M_\mathrm{d0} = 0.01\,M_*$
    (model~\texttt{DP-M}).
    Top left: the angles that the angular momentum vectors $\boldsymbol{D}$,
    $\boldsymbol{P}$ and $\boldsymbol{J}_\mathrm{dp}$ form with the $z$ axis
    (the initial direction of $\boldsymbol{D}$), as well as the angle between
    $\boldsymbol{D}$ and $\boldsymbol{P}$.
    Top right: the projections of the angular momentum unit vectors onto the
    $x$--$y$ plane.
    Bottom left: the characteristic precession frequency.
    Bottom right: the magnitudes of the angular momentum vectors.
    In the left-hand panels, the initial $0.1$\,Myr of the evolution is
    displayed at a higher resolution.
    \label{fig:DP-M}}
\end{figure*}
Figure~\ref{fig:DP-M} shows the evolution of such a system for the case
(model~\texttt{DP-M}) where a Jupiter-mass planet on a misaligned orbit
($\psi_\mathrm{p0} = 60^\circ$) torques an inner disk of initial mass
$M_\mathrm{d0} = 0.01\,M_*$ (corresponding to $M_\mathrm{t0} = 0.1\,M_*$, i.e.,
to $t_0 = 0$ when $M_* = M_\sun$; see Equation~(\ref{eq:deplete})).
The top left panel exhibits the angles $\psi_\mathrm{d}$ and $\psi_\mathrm{p}$
(blue: inner disk; red: planet) as a function of time.
In this and the subsequent figures, we show results for a total duration of
$10$\,Myr.
This is long enough in comparison with $\tau$ to capture the secular evolution
of the system, which is driven by the mass depletion in the inner disk.
To capture the details of the oscillatory behavior associated with the
precession of the individual angular momentum vectors ($\boldsymbol{D}$ and
$\boldsymbol{P}$) about the total angular momentum vector
$\boldsymbol{J}_\mathrm{dp} = \boldsymbol{D} + \boldsymbol{P}$ (subscript
j)---which takes place on the shorter timescale $\tau_\mathrm{dp}$
($\simeq 9\times 10^3$\,yr at $t = t_0$)---we display the initial $0.1$\,Myr in
the top left panel using a higher time resolution and, in addition, show the
projected trajectories of the unit vectors $\hat{\boldsymbol{D}}$,
$\hat{\boldsymbol{P}}$, and $\hat{\boldsymbol{J}}_\mathrm{dp}$ in the $x$--$y$
plane during this time interval in the top right panel.
Given that $0.1\,{\rm Myr} \ll \tau$, the vectors $\hat{\boldsymbol{D}}$ and
$\hat{\boldsymbol{P}}$ execute a circular motion about
$\hat{\boldsymbol{J}}_\mathrm{dp}$ with virtually constant inclinations with
respect to the latter vector (given by the angles $\theta_\mathrm{jd}$ and
$\theta_\mathrm{jp}$, respectively), and the orientation of
$\hat{\boldsymbol{J}}_\mathrm{dp}$ with respect to the $z$ axis (given by the
angle $\psi_\mathrm{j}$) also remains essentially unchanged.
(The projection of $\hat{\boldsymbol{J}}_\mathrm{dp}$  on the $x$--$y$ plane is
displaced from the center along the $y$ axis, reflecting the fact that the
planet's initial line of nodes coincides with the $x$ axis.)
As the vectors $\hat{\boldsymbol{D}}$ and $\hat{\boldsymbol{P}}$ precess about
$\hat{\boldsymbol{J}}_\mathrm{dp}$, the angles $\psi_\mathrm{d}$ and
$\psi_\mathrm{p}$ oscillate in the ranges
$|\psi_\mathrm{j} - \theta_\mathrm{jd}| \leq \psi_\mathrm{d} \leq
\psi_\mathrm{j} + \theta_\mathrm{jd}$ and
$|\psi_\mathrm{j} - \theta_\mathrm{jp}| \leq \psi_\mathrm{p} \leq
\psi_\mathrm{j} + \theta_\mathrm{jp}$, respectively.

\begin{figure}
  \begin{center}
    \includegraphics{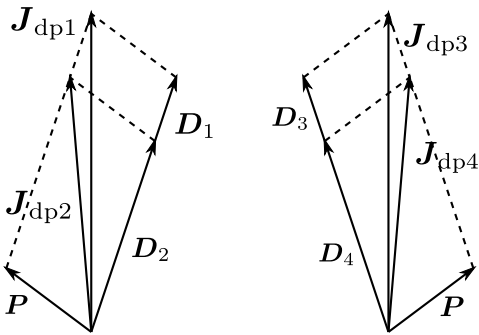}
  \end{center}
  \caption{
    Schematic sketch of the change in the total angular momentum vector
    $\boldsymbol{J}_\mathrm{dp}$ that is induced by mass depletion from the disk
    in the limit where the precession period $\tau_{\rm dp}$ is much shorter
    than the characteristic depletion time $\tau$.
    The two depicted configurations are separated by $0.5\,\tau_\mathrm{dp}$.
  \label{fig:vectors}}
\end{figure}
A notable feature of the evolution of this system on a timescale $\gtrsim \tau$
is the increase in the angle $\psi_\mathrm{d}$ (blue line in the top left
panel)---indicating progressive misalignment of the disk with respect to its
initial orientation---as the magnitude of the angular momentum $\boldsymbol{D}$
decreases with the loss of mass from the disk (blue line in the bottom right
panel).
At the same time, the orbital plane of the planet (red line in the top left
panel) tends toward alignment with $\boldsymbol{J}_\mathrm{dp}$.
The magenta lines in the top left and bottom right panels indicate that the
orientation of the vector $\boldsymbol{J}_\mathrm{dp}$ remains fixed even as its
magnitude decreases (on a timescale $\gtrsim \tau$) on account of the decrease
in the magnitude of $\boldsymbol{D}$.
As we demonstrate analytically in Appendix~\ref{app:Jdp}, the constancy of
$\psi_\mathrm{j}$ is a consequence of the inequality
$\tau_\mathrm{dp} \ll \tau$.

To better understand the evolution of the disk and planet orientations, we
consider the (small) variations in $\boldsymbol{D}$ and
$\boldsymbol{J}_\mathrm{dp}$ that are induced by mass depletion over a small
fraction of the precession period.
On the left-hand side of Figure~\ref{fig:vectors} we show a schematic sketch of
the orientations of the vectors $\boldsymbol{D}$, $\boldsymbol{P}$, and
$\boldsymbol{J}_\mathrm{dp}$ at some given time (denoted by the subscript 1) and
a short time later (subscript 2).
During that time interval the vector $\boldsymbol{J}_\mathrm{dp}$ tilts slightly
to the left, and as a result it moves away from $\boldsymbol{D}$ and closer to
$\boldsymbol{P}$.
The sketch on the right-hand side of Figure~\ref{fig:vectors} demonstrates that,
if we were to consider the same evolution a half-cycle later, the same
conclusion would be reached: in this case the vector
$\boldsymbol{J}_{\mathrm{dp}3}$ moves slightly to the right (to become
$\boldsymbol{J}_{\mathrm{dp}4}$), with the angle between
$\boldsymbol{J}_\mathrm{dp}$ and $\boldsymbol{D}$ again increasing even as the
angle between $\boldsymbol{J}_\mathrm{dp}$ and $\boldsymbol{P}$ decreases.
The angles between the total angular momentum vector and the vectors
$\boldsymbol{D}$ and $\boldsymbol{P}$ are thus seen to undergo a systematic,
secular variation.
The sketch in Figure~\ref{fig:vectors} also indicates that the vector
$\boldsymbol{J}_\mathrm{dp}$ undergoes an oscillation over each precession
cycle.
However, when $\tau_\mathrm{dp} \ll \tau$ and the fractional decrease in
$M_\mathrm{d}$ over a precession period remains $\ll 1$, the amplitude of the
oscillation is very small and $\boldsymbol{J}_\mathrm{dp}$ practically maintains
its initial direction (see Appendix~\ref{app:Jdp} for a formal demonstration of
this result).
In the limit where the disk mass becomes highly depleted and $D \to 0$,
$\boldsymbol{J}_\mathrm{dp} \to \boldsymbol{P}$, i.e., the planet aligns with
the initial direction of $\boldsymbol{J}_\mathrm{dp}$
($\theta_\mathrm{jp} \to 0$ and $\psi_\mathrm{p} \to \psi_\mathrm{j}$).
The disk angular momentum vector then precesses about $\boldsymbol{P}$, with its
orientation angle $\psi_\mathrm{d}$ (blue line in top left panel of
Figure~\ref{fig:DP-M}) oscillating between
$|\psi_\mathrm{p} - \theta_\mathrm{dp}|$ and
$\psi_\mathrm{p} + \theta_\mathrm{dp}$.\footnote{
The angle $\theta_\mathrm{dp}$ between $\boldsymbol{D}$ and $\boldsymbol{P}$
(cyan line in the top left panel of Figure~\ref{fig:DP-M}) remains constant
because there are no torques that can modify it.}
Note that the precession frequency is also affected by the disk's mass depletion
and decreases with time (see Equation~(\ref{eq:omega_dp})); the time evolution
of $\Omega_{\rm dp}$ is shown in the bottom left panel of Figure~\ref{fig:DP-M}.

\begin{figure*}
  \includegraphics[width=\textwidth]{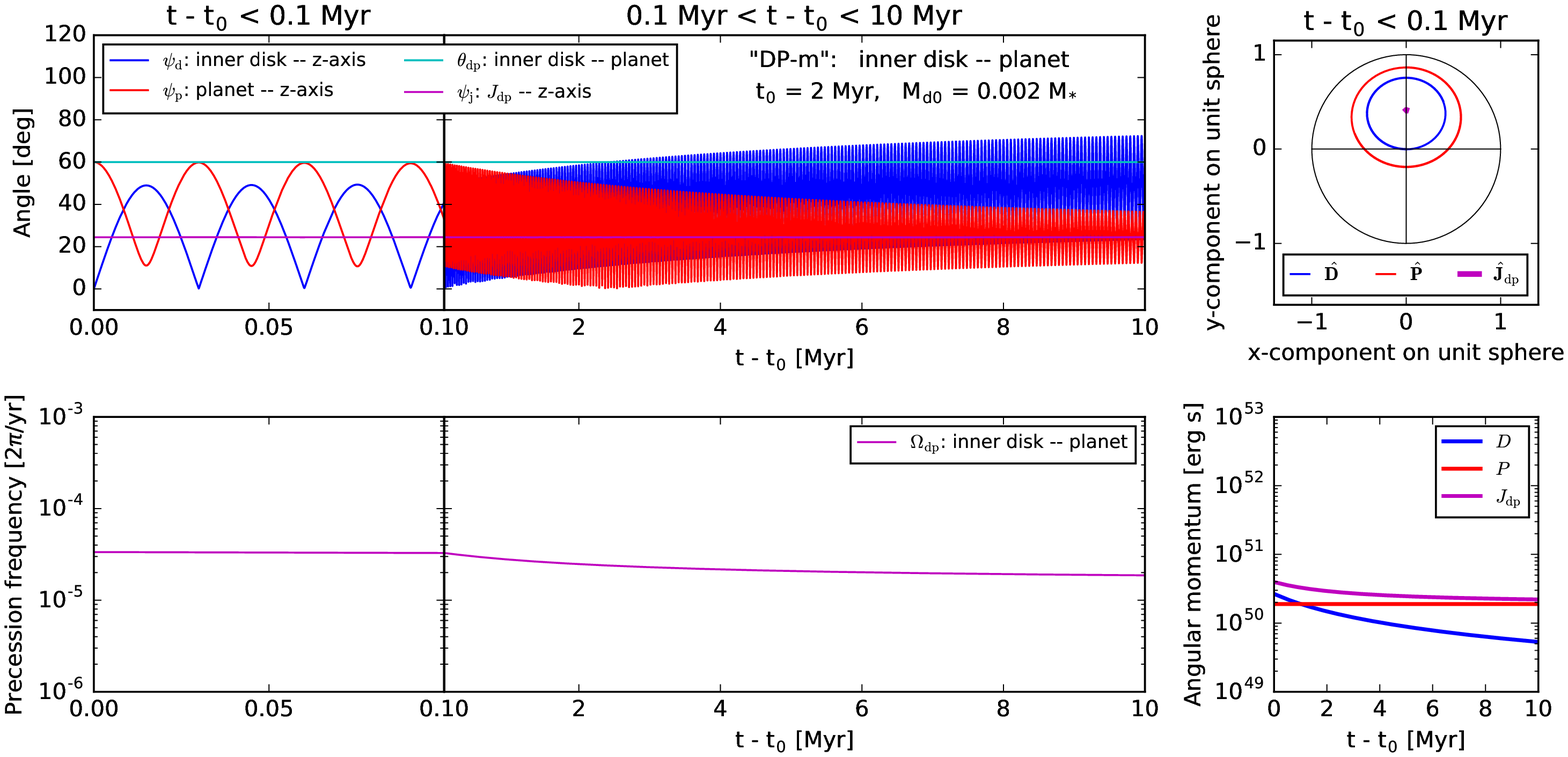}
  \caption{
    Same as Figure~\ref{fig:DP-M}, except that $M_\mathrm{d0} = 0.02\,M_*$
    (model~\texttt{DP-m}).
  \label{fig:DP-m}}
\end{figure*}
Figure~\ref{fig:DP-m} shows the evolution of a similar
system---model~\texttt{DP-m}---in which the inner disk has a lower initial mass,
$M_\mathrm{d0} = 0.002\,M_*$ (corresponding to $M_\mathrm{t0} = 0.02\,M_*$,
i.e., to $t_0=2$\,Myr when $M_*=M_\sun$; see Equation~(\ref{eq:deplete})).
The initial oscillation frequency in this case is lower than in model
\texttt{DP-M}, as expected from Equation~(\ref{eq:omega_dp}), but it attains the
same asymptotic value (bottom left panel), corresponding to the limit
$J_\mathrm{dp} \to P$ in which $\Omega_\mathrm{dp}$ becomes independent of
$M_\mathrm{d}$.
The initial value of $J_\mathrm{dp}/D$ is higher in the present model than in
the model considered in Figure~\ref{fig:DP-M} ($\simeq 1.5$ vs. $\simeq 1.1$;
see Equations~(\ref{eq:P}) and~(\ref{eq:D})), which results in a higher value of
$\psi_\mathrm{j}$ (and, correspondingly, a higher initial value of
$\theta_\mathrm{jd}$ and lower initial value of $\theta_\mathrm{jp}$).
The higher value of $\psi_\mathrm{j}$ is the reason why the oscillation
amplitude of $\psi_\mathrm{d}$ and the initial oscillation amplitude of
$\psi_\mathrm{p}$ (top left panel) are larger in this case.
The higher value of $J_\mathrm{dp}/D_0$ in Figure~\ref{fig:DP-m} also accounts
for the differences in the projection map shown in the top right panel (a larger
$y$ value for the projection of $\hat{\boldsymbol{J}}_\mathrm{dp}$, a larger
area encircled by the projection of $\hat{\boldsymbol{D}}$, and a smaller area
encircled by the projection of $\hat{\boldsymbol{P}}$).

\begin{figure*}
  \includegraphics[width=\textwidth]{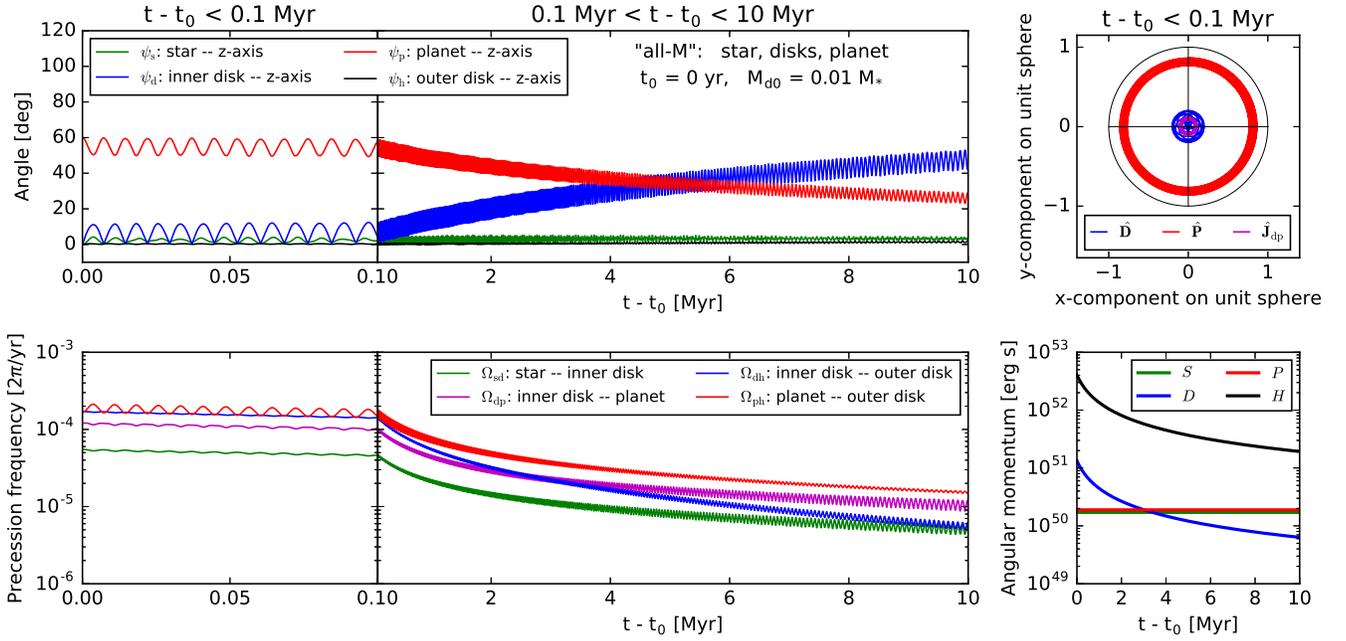}
  \caption{
    Time evolution of the full system (star, inner disk, planet, outer disk) for
    an initial inner disk mass $M_\mathrm{d0} = 0.01\,M_*$ and initial total
    disk mass $M_\mathrm{t0} = 0.1\,M_*$ (model~\texttt{all-M}).
    Panel arrangement is the same as in Figure~\ref{fig:DP-M}, although the
    details of the displayed quantities---which are specified in each panel and
    now also include the angular momenta of the star ($\boldsymbol{S}$) and the
    outer disk ($\boldsymbol{H}$)---are different.
  \label{fig:all-M}}
\end{figure*}
\begin{figure*}
  \includegraphics[width=\textwidth]{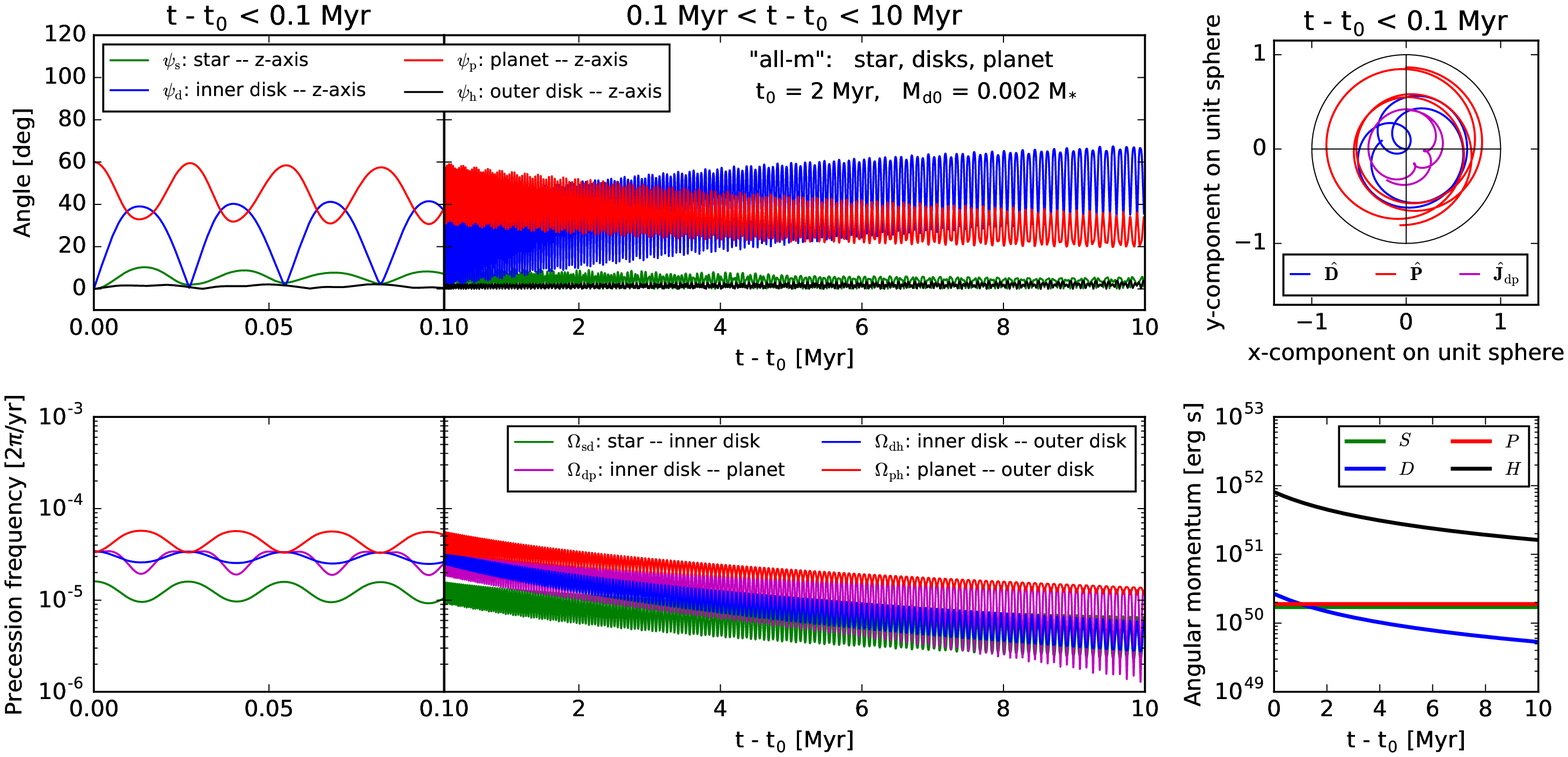}
  \caption{
    Same as Figure~\ref{fig:all-M}, except that $M_\mathrm{d0} = 0.002\,M_*$ and
    $M_\mathrm{t0} = 0.02\,M_*$ (model~\texttt{all-m}).
  \label{fig:all-m}}
\end{figure*}
We now consider the full system for two values of the total disk mass:
$M_\mathrm{t0} = 0.1\,M_*$ (model~\texttt{all-M}, corresponding to $t_0 = 0$;
Figure~\ref{fig:all-M}) and $M_\mathrm{t0} = 0.02\,M_*$ (model~\texttt{all-m},
corresponding to $t_0 = 2$\,Myr; Figure~\ref{fig:all-m}), assuming that both
parts of the disk lose mass according to the relation given by
Equation~(\ref{eq:deplete}).
The inner disks in these two cases correspond, respectively, to the disk masses
adopted in model~\texttt{DP-M} (Figure~\ref{fig:DP-M}) and model~\texttt{DP-m}
(Figure~\ref{fig:DP-m}).
The merit of first considering the simpler systems described by the latter
models becomes apparent from a comparison between the respective figures.
It is seen that the basic behavior of model~\texttt{all-M} is similar to that of
model~\texttt{DP-M}, and that the main differences between model~\texttt{all-M}
and model~\texttt{all-m} are captured by the way in which model~\texttt{DP-m} is
distinct from model~\texttt{DP-M}.
The physical basis for this correspondence is the centrality of the torque
exerted on the inner disk by the planet.
According to Equation~(\ref{eq:precession}), the relative magnitudes of the
torques acting on the disk at sufficiently late times (after $D$ becomes smaller
than the angular momentum of each of the other system components) are reflected
in the magnitudes of the corresponding precession frequencies.
The dominance of the planet's contribution can thus be inferred from the plots
in the bottom left panels of Figures~\ref{fig:all-M} and~\ref{fig:all-m}, which
show that, after the contribution of $D$ becomes unimportant (bottom right
panels), the precession frequency induced by the planet exceeds those induced by
the outer disk and by the star.\footnote{
The star--planet and star--outer-disk precession frequencies
($\Omega_\mathrm{sp}$ and~$\Omega_\mathrm{sh}$; see
Equations~(\ref{eq:omega_sp}) and~(\ref{eq:omega_sh})) are not shown in these
figures because they are too low to fit in the plotted range.}

While the basic disk misalignment mechanism is the same as in the
planet--inner-disk system, the detailed behavior of the full system is
understandably more complex.
One difference that is apparent from a comparison of the left-hand panels in
Figures~\ref{fig:all-M} and~\ref{fig:DP-M} is the higher oscillation frequency
of $\psi_\mathrm{p}$ and $\psi_\mathrm{d}$ in the full model (with the same
frequency also seen in the timeline of $\psi_\mathrm{s}$).
In this case the planet--outer-disk  precession frequency $\Omega_\mathrm{ph}$
(Equation~(\ref{eq:omega_ph})) and the inner-disk--outer-disk precession
frequency $\Omega_\mathrm{dh}$ (Equation~(\ref{eq:omega_dh})) are initially
comparable and larger than $\Omega_\mathrm{dp}$, and $\Omega_\mathrm{ph}$
remains the dominant frequency throughout the system's evolution.
The fact that the outer disk imposes a precession on both $\boldsymbol{P}$ and
$\boldsymbol{D}$ has the effect of weakening the interaction between the planet
and the inner disk, which slows down the disk misalignment process.
Another difference is revealed by a comparison of the top right panels: in the
full system, $\hat{\boldsymbol{J}}_\mathrm{dp}$ precesses on account of the
torque induced by the outer disk, so it no longer corresponds to just a single
point in the $x$--$y$ plane.
This, in turn, increases the sizes of the regions traced in this plane by
$\hat{\boldsymbol{D}}$ and $\hat{\boldsymbol{P}}$.
The behavior of the lower-$M_\mathrm{t0}$ model shown in Figure~\ref{fig:all-m}
is also more involved.
In this case, in addition to the strong oscillations of the angles $\psi_i$
already manifested in Figure~\ref{fig:DP-m}, the different precession
frequencies $\Omega_{ik}$ also exhibit large-amplitude oscillations, reflecting
their dependence on the angles $\theta_{ik}$ between the angular momentum
vectors.
In both of the full-system models, the strongest influence on the star is
produced by its interaction with the inner disk, but the resulting precession
frequency ($\Omega_\mathrm{sd}$) remains low.
Therefore, the stellar angular momentum vector essentially retains its original
orientation, which implies that the angle $\psi_\mathrm{d}$ is a good proxy for
the angle between the primordial stellar spin and the orbit of any planet that
eventually forms in the inner disk.

\begin{figure}
  \includegraphics[width=\columnwidth]{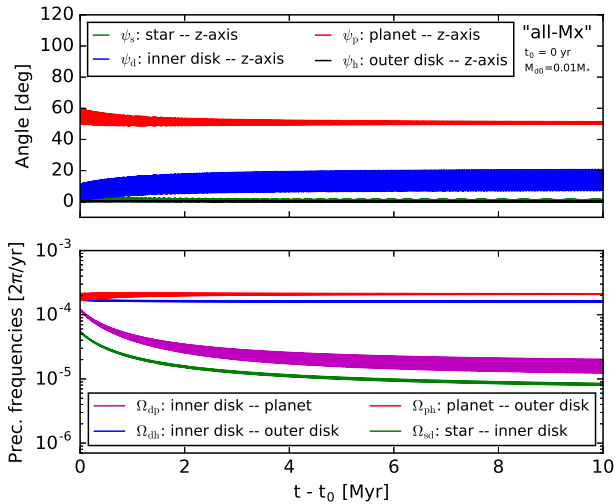}
  \caption{
    Time evolution of the full system in the limit where only the inner disk
    undergoes mass depletion and the mass of the outer disk remains unchanged,
    for the same initial conditions as in Figure~\ref{fig:all-M}
    (model~\texttt{all-Mx}).
    The top and bottom panels correspond, respectively, to the top left and
    bottom left panels of Figure~\ref{fig:all-M}, but in this case the initial
    $0.1$\,Myr of the evolution are not displayed at a higher resolution.
  \label{fig:all-Mx}}
\end{figure}
\begin{figure}
  \includegraphics[width=\columnwidth]{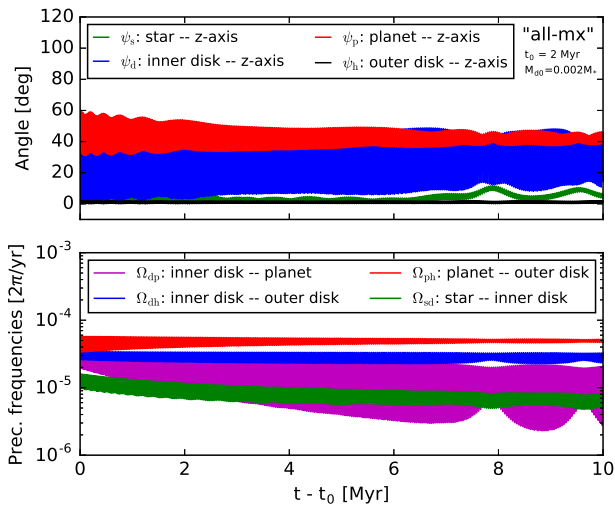}
  \caption{
    Same as Figure~\ref{fig:all-Mx}, but for the initial conditions of
    Figure~\ref{fig:all-m} (model~\texttt{all-mx}).
  \label{fig:all-mx}}
\end{figure}
We repeated the calculations shown in Figures~\ref{fig:all-M}
and~\ref{fig:all-m} under the assumption that only the inner disk loses mass
while $M_\mathrm{h}$ remains constant (models~\texttt{all-Mx}
and~\texttt{all-mx}; Figures~\ref{fig:all-Mx} and~\ref{fig:all-mx},
respectively).
At the start of the evolution, the frequencies $\Omega_\mathrm{ph}$ and
$\Omega_\mathrm{dh}$ are $\propto$$M_\mathrm{h}$, whereas $\Omega_\mathrm{dp}$
scales linearly (or, in the case of the lower-$M_\mathrm{d0}$ model, close to
linearly) with $M_\mathrm{d}$ (see Appendix~\ref{app:torques}).
In the cases considered in Figures~\ref{fig:all-M} and~\ref{fig:all-m} all these
frequencies decrease with time, so the relative magnitude of
$\Omega_\mathrm{dp}$ remains comparatively large throughout the evolution.
In contrast, in the cases shown in Figures~\ref{fig:all-Mx} and~\ref{fig:all-mx}
the frequencies $\Omega_\mathrm{ph}$ and $\Omega_\mathrm{dh}$ remain constant
and only $\Omega_\mathrm{dp}$ decreases with time.
As the difference between $\Omega_\mathrm{dp}$ and the other two frequencies
starts to grow, the inner disk misalignment process is aborted, and thereafter
the mean values of $\psi_\mathrm{d}$ and $\psi_\mathrm{p}$ remain constant.
This behavior is consistent with our conclusion about the central role that the
torque exerted by the planet plays in misaligning the inner disk: when the fast
precession that the outer disk induces in the orbital motions of both the planet
and the inner disk comes to dominate the system dynamics, the direct coupling
between the planet and the inner disk is effectively broken and the misalignment
process is halted.
Note, however, from Figure~\ref{fig:all-mx} that, even in this case, the angle
$\psi_\mathrm{d}$ can attain a high value (as part of a large-amplitude
oscillation) when $M_\mathrm{t0}$ is small.

\begin{figure*}
  \includegraphics[width=\textwidth]{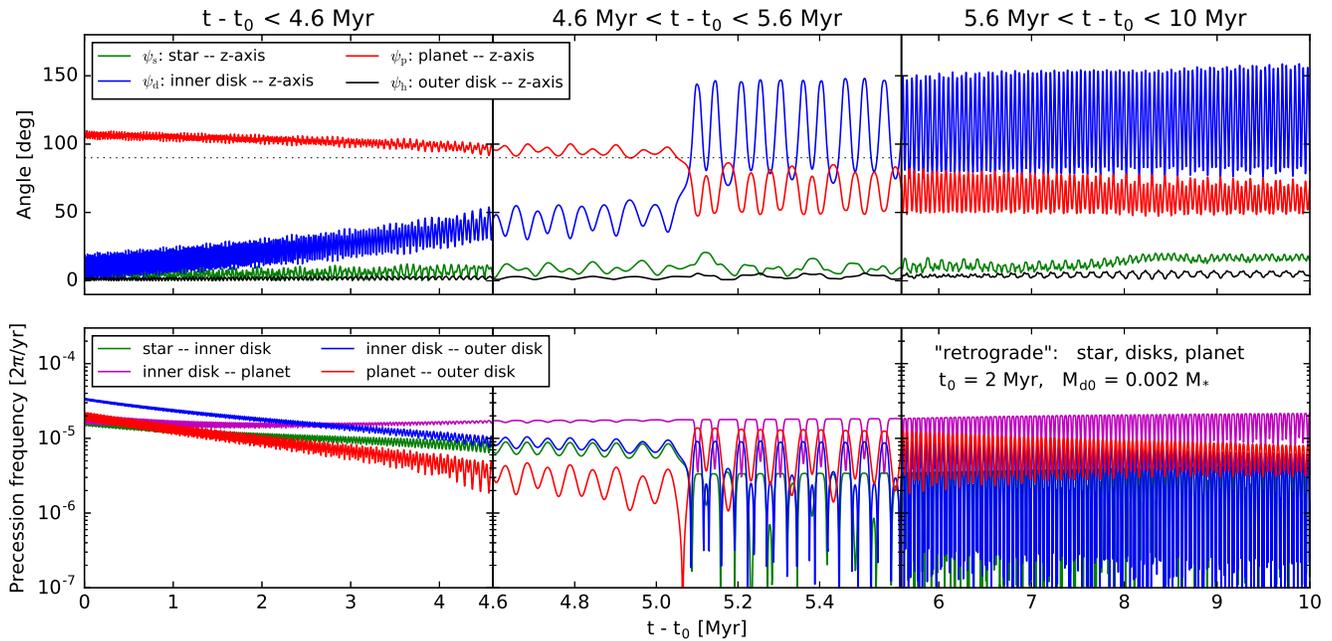}
  \caption{
    Time evolution with the same initial conditions as in
    Figure~\ref{fig:all-m}, except that the planet is initially on a retrograde
    orbit ($\psi_{\mathrm{p}0}$ is changed from $60^\circ$ to $110^\circ$;
    model~\texttt{retrograde}).
    The display format is the same as in Figure~\ref{fig:all-Mx}, but in this
    case the panels also show a zoomed-in version of the evolution around the
    time of the jumps in $\psi_\mathrm{p}$ and $\psi_\mathrm{d}$.
    The dashed line in the top panel marks the transition between prograde and
    retrograde orientations ($90^\circ$).
  \label{fig:retrograde}}
\end{figure*}
To determine whether the proposed misalignment mechanism can also account for
disks (and, eventually, planets) on retrograde orbits, we consider a system in
which the companion planet is placed on such an orbit
(model~\texttt{retrograde}, which is the same as model~\texttt{all-m} except
that $\psi_{\mathrm{p}0}$ is changed from $60^\circ$ to $110^\circ$).
As Figure~\ref{fig:retrograde} demonstrates, the disk in this case evolves to a
retrograde configuration ($\psi_\mathrm{d} > 90^\circ$) at late times even as
the planet's orbit reverts to prograde motion.
A noteworthy feature of the plotted orbital evolution (shown in the
high-resolution portion of the figure) is the rapid increase in the value of
$\psi_\mathrm{d}$ (which is an adequate proxy for $\theta_\mathrm{sd}$ also in
this case)---and corresponding fast decrease in the value of
$\psi_\mathrm{p}$---that occurs when the planet's orbit transitions from a
retrograde to a prograde orientation.
This behavior can be traced to the fact that $\cos{\theta_\mathrm{ph}}$ vanishes
at essentially the same time that $\psi_\mathrm{p}$ crosses $90^\circ$ because
the outer disk (which dominates the total angular momentum) remains well aligned
with the $z$ axis.
This, in turn, implies (see Equation~(\ref{eq:omega_ph})) that, at the time of
the retrograde-to-prograde transition, the planet becomes dynamically decoupled
from the outer disk and only retains a coupling to the inner disk.
Its evolution is, however, different from that of a ``reduced'' system, in which
only the planet and the inner disk interact, because the inner disk remains
dynamically ``tethered'' to the outer disk ($\theta_\mathrm{dh}\ne 90^\circ$).
As we verified by an explicit calculation, the evolution of the reduced system
remains smooth when $\psi_\mathrm{p}$ crosses $90^\circ$.
The jump in $\psi_\mathrm{p}$ exhibited by the full system leads to a
significant increase in the value of $\cos{\theta_\mathrm{ph}}$ and hence of
$\Omega_\mathrm{ph}$, which, in turn, restores (and even enhances) the planet's
coupling to the outer disk after its transition to retrograde motion (see bottom
panel of Figure~\ref{fig:retrograde}).
The maximum value attained by $\theta_\mathrm{sd}$ in this example is
$\simeq 172^\circ$, which, just as in the prograde case shown in
Figure~\ref{fig:all-m}, exceeds the initial misalignment angle of the planetary
orbit (albeit to a much larger extent in this case).
It is, however, worth noting that not all model systems in which the planet is
initially on a retrograde orbit give rise to a retrograde inner disk at the end
of the prescribed evolution time; in particular, we found that the outcome of
the simulated evolution (which depends on whether $\psi_\mathrm{p}$ drops below
$90^\circ$) is sensitive to the value of the initial planetary misalignment
angle $\psi_{\mathrm{p}0}$ (keeping all other model parameters unchanged).

In concluding this section it is instructive to compare the results obtained
for our model with those found for the model originally proposed by
\citet{Batygin12} (see Section~\ref{sec:intro} for references to additional work
on that model).
We introduced our proposed scenario as a variant of the latter model, with a
close-by giant planet taking the place of a distant stellar companion.
In the original proposal the disk misalignment was attributed to the
precessional motion that is induced by the torque that the binary companion
exerts on the disk.
In this picture the spin--orbit angle oscillates (on a timescale $\sim$1\,Myr
for typical parameters) between $0^\circ$ and roughly twice the binary orbital
inclination, so it can be large if observed at the ``right'' time.
Our model retains this feature of the earlier proposal, particularly in cases
where the companion planet is placed on a high-inclination orbit after the disk
has already lost much of its initial mass, but it also exhibits a novel feature
that gives rise to a secular (rather than oscillatory) change in the spin--orbit
angle (which can potentially lead to a substantial increase in this angle).
This new behavior represents an ``exchange of orientations'' between the planet
and the inner disk that is driven by the mass loss from the inner disk and
corresponds to a decrease of the inner disk's angular momentum from a value
higher than that of the planet to a lower value (with the two remaining within
an order of magnitude of each other for representative parameters).
This behavior is not found in a binary system because of the large mismatch
between the angular momenta of the companion and the disk in that case (and, in
fact, it is also suppressed in the case of a planetary companion when the mass
of the outer disk is not depleted). 

As we already noted in Section~\ref{subsec:assumptions}, \citet{BatyginAdams13}
suggested that the disk misalignment in a binary system can be significantly
increased due to a resonance between the star--disk and binary--disk precession
frequencies.
(We can use Equations~(\ref{eq:omega_sd}) and~(\ref{eq:omega_dp}), respectively,
to evaluate these frequencies, plugging in values for the outer disk radius,
companion orbital radius, and companion mass that are appropriate for the binary
case.)
\citet{Lai14} clarified the effect of this resonance and emphasized that, for
plausible system parameters, it can be expected to be crossed as the disk
becomes depleted of mass.
However, for the planetary-companion systems considered in this paper the ratio
$|\Omega_\mathrm{sd}/\Omega_\mathrm{dp}|$ remains $< 1$ throughout the
evolution, so no such resonance is encountered in this case.
In both of these systems $\Omega_\mathrm{sd}$ is initially
$\propto M_\mathrm{d}$, so it decreases during the early evolution.
The same scaling also characterizes $\Omega_\mathrm{dp}$ in the planetary case,
which explains why the corresponding curves do not cross.
In contrast, in the binary case (for which the sum of the disk and companion
angular momenta is dominated by the companion's contribution) the frequency
$\Omega_\mathrm{dp}$ does not scale with the disk mass and it thus remains
nearly constant, which makes it possible for the corresponding curves to cross
(see Figure~\ref{fig:binary} in Appendix~\ref{app:resonance}).
Since our formalism also encompasses the binary case, we examined one such
system (model~\texttt{binary})---using the parameters adopted in figure~3 of
\citet{Lai14}---for comparison with the results of that work.
Our findings are presented in Appendix~\ref{app:resonance}.

\section{Discussion}
  \label{sec:discussion}
  
The model considered in this paper represents a variant of the primordial disk
misalignment scenario of \citet{Batygin12} in which the companion is a nearby
planet rather than a distant star and only the inner region of the
protoplanetary disk (interior to the planet's orbit) becomes inclined.
In this section we assess whether this model provides a viable framework for
interpreting the relevant observations.

The first---and most basic---question that needs to be addressed is whether the
proposed misalignment mechanism is compatible with the broad range of apparent
spin--orbit angles indicated by the data.
In Section~\ref{sec:results} we showed that the spin--orbit angle
$\theta_\mathrm{sd}$ can deviate from its initial  value of $0^\circ$ either
because of the precessional motion that is induced by the planet's torque on the
disk or on account of the secular variation that is driven by the mass depletion
process.
In the ``reduced'' disk--planet model considered in Figures~\ref{fig:DP-M}
and~\ref{fig:DP-m}, for which the angle $\psi_\mathrm{d}$ is taken as a proxy
for the intrinsic spin--orbit angle, the latter mechanism increases
$\theta_\mathrm{sd}$ to $\sim$$45^\circ$--$50^\circ$ on a timescale of $10$\,Myr
for an initial planetary inclination $\psi_\mathrm{p0} = 60^\circ$.
The maximum disk misalignment is, however, increased above this value by the
precessional oscillation, whose amplitude is higher the lower the initial mass
of the disk.
Based on the heuristic discussion given in connection with
Figure~\ref{fig:vectors}, the maximum possible value of $\psi_\mathrm{d}$
(corresponding to the limit $J_\mathrm{dp} \to P$) is given by
\begin{equation}
  \label{eq:psi_max}
  \psi_\mathrm{d,max} = \arccos\frac{D_0 + P\cos\psi_\mathrm{p0}}
  {(D_0^2 + P^2 + 2D_0P\cos\psi_\mathrm{p0})^{1/2}} + \psi_\mathrm{p0}\,.
\end{equation}
For the parameters of Figure~\ref{fig:DP-m},
$\psi_\mathrm{d,max} \approx 84.5^\circ$, which can be compared with the actual
maximum value ($\simeq 72^\circ$) attained over the course of the $10$-Myr
evolution depicted in this figure.\footnote{
The intrinsic spin--orbit angle is not directly measurable, so its value must be
inferred from that of the apparent (projected) misalignment angle $\lambda$
\citep{FabryckyWinn09}.
In the special case of a planet whose orbital plane contains the line of
sight---an excellent approximation for planets observed by the transits
method---the apparent obliquity cannot exceed the associated intrinsic
misalignment angle (i.e., $\lambda \le \theta_\mathrm{sd}$).}
Although the behavior of the full system (which includes also the outer disk and
the star) is more complicated, we found (see Figures~\ref{fig:all-M}
and~\ref{fig:all-m}) that, if the outer disk also loses mass, the maximum value
attained by $\theta_{\rm sd}$ ($\simeq 67^\circ$) is not much smaller than in
the simplified model.
Note that in the original primordial-misalignment scenario the maximum value of
$\theta_\mathrm{sd}$ ($\simeq 2\,\psi_\mathrm{p0}$) would have been considerably
higher ($\simeq 120^\circ$) for the parameters employed in our example.
However, as indicated by Equation~(\ref{eq:psi_max}), the maximum value
predicted by  our model depends on the ratio $P/D_0$ and can in principle exceed
the binary-companion limit if $D_0$ is small and $P$ is sufficiently
large.\footnote{
$D_0$, the magnitude of the initial angular momentum of the inner disk, cannot
be much smaller than the value adopted in models~\texttt{DP-m}
and~\texttt{all-m} in view of the minimum value of $M_\mathrm{d0}$ that is
needed to account for the observed misaligned planets in the
primordial-disk-misalignment scenario (and also for the no-longer-present HJ in
the SHJ picture).}
Repeating the calculations shown in Figure~\ref{fig:all-m} for higher values of 
$M_\mathrm{p}$, we found that the maximum value of $\theta_\mathrm{sd}$ is 
$\sim$$89^\circ$, $104^\circ$ and~$125^\circ$ when $M_\mathrm{p}/M_\mathrm{J}$
increases from~1 to~2, 3, and~4, respectively.
These results further demonstrate that the disk can be tilted to a retrograde
configuration even when $\psi_\mathrm{p0} < 90^\circ$ if the planet is
sufficiently massive, although a retrograde disk orientation can also be
attained (including in the case of $M_\mathrm{p} \lesssim M_\mathrm{J}$) if the
planet's orbit is initially retrograde (see Figure~\ref{fig:retrograde}).
A low initial value of the disk angular momentum $D$ arises naturally in the
leading scenarios for placing planets in inclined orbits, which favor
comparatively low disk masses (see Section~\ref{sec:intro}).
The distribution of $\psi_\mathrm{p0}$ as well as those of the occurrence rate,
mass, and orbital radius of planets on inclined orbits are required for
determining the predicted distribution of primordial inner-disk misalignment
angles in this scenario, for comparison with observations.\footnote{
\citet{MatsakosKonigl15} were able to reproduce the observed obliquity
distributions of HJs around G and F stars within the framework of the SHJ model
under the assumption that the intrinsic spin--orbit angle has a random
distribution (corresponding to a flat distribution of $\lambda$; see
\citealt{FabryckyWinn09}).}
However, this information, as well as data on the relevant values of
$M_\mathrm{d0}$, are not yet available, so our results for $\theta_\mathrm{sd}$
are only a first step (a proof of concept) toward validating this interpretation
of the measured planet obliquities.

Our proposed misalignment mechanism is most effective when the disk mass within
the planetary orbit drops to $\sim$$M_\mathrm{p}$.
In the example demonstrating this fact (Figure~\ref{fig:all-m}),
$M_\mathrm{d0} \approx 2\,M_\mathrm{J}$.
In the primordial disk misalignment scenario, $M_\mathrm{d0}$ includes the mass
that would eventually be detected in the form of an HJ (or a lower-mass planet)
moving around the central star on a misaligned orbit.
Furthermore, if the ingestion of an HJ on a misaligned orbit is as ubiquitous as
inferred in the SHJ picture, that mass, too, must be included in the tally.
These requirements are consistent with the fact that the typical disk
misalignment time in our model (a few Myr) is comparable to the expected
giant-planet formation time, but this similarity also raises the question of
whether the torque exerted by the initially misaligned planet has the same
effect on the gaseous inner disk and on a giant planet embedded within it.
This question was considered by several authors in the context of a binary
companion \citep[e.g.,][]{Xiang-GruessPapaloizou14, PicognaMarzari15,
Martin+16}.
A useful gauge of the outcome of this dynamical interaction is the ratio of the
precession frequency induced in the embedded planet (which we label
$\Omega_\mathrm{pp}$) to $\Omega_\mathrm{dp}$ \citep{PicognaMarzari15}.
We derive an expression for $\Omega_\mathrm{pp}$ by approximating the inclined
and embedded planets as two rings with radii $a$ and $a_1 < a$, respectively
(see Appendix~\ref{app:torques}), and evaluate $\Omega_\mathrm{dp}$ under the
assumption that the disk mass has been sufficiently depleted for the planetary
contribution ($P$) to dominate $J_\mathrm{dp}$.
This leads to
$\Omega_\mathrm{pp}/\Omega_\mathrm{dp} \simeq 2\,(a_1/r_\mathrm{d,out})^{3/2}$,
which is the same as the estimate obtained by \citet{PicognaMarzari15} for a
binary system.
In the latter case, this ratio is small ($\lesssim 0.1$) for typical parameters,
implying that the embedded planet cannot keep up with the disk precession and
hence that its orbit develops a significant tilt with respect to the disk's
plane.
However, when the companion is a planet, the above ratio equals $(a_1/a)^{3/2}$
and may be considerably larger ($\lesssim 1$), which suggests that the embedded
planet can remain coupled to the disk in this case.

A key prediction of our proposed scenario---which distinguishes it from the
original \citet{Batygin12} proposal---is that there would in general be a
difference in the obliquity properties of ``nearby'' and ``distant'' planets,
corresponding to the different orientations attained, respectively, by the inner
and outer disks.
This prediction is qualitatively consistent with the finding of \citet{LiWinn16}
that the good spin--orbit alignment inferred in cool stars from an analysis of
rotational photometric modulations in \textit{Kepler} sources \citep{Mazeh+15}
becomes weaker (with the inferred orientations possibly tending toward a nearly
random distribution) at large orbital periods
($P_\mathrm{orb} \gtrsim 10^2\,$days).
The interpretation of these results in our picture is that the outer planets
remain aligned with the original stellar-spin direction, whereas the inner
planets---and, according to the SHJ model, also the stellar spin in $\sim$50\%
of sources---assume the orientation of the misaligned inner disk (which samples
a broad range of angles with respect to the initial spin direction).
Further observations and analysis are required to corroborate and refine these
findings so that they can be used to place tighter constrains on the models.

The result reported by \citet{LiWinn16} is seemingly at odds with another set of
observational findings---the discovery that the orbital planes of debris disks
(on scales $\gtrsim 10^2\,$au) are by and large well aligned with the spin axis
of the central star \citep{Watson+11, Greaves+14}.
This inferred alignment also seemingly rules out any interpretation of the
obliquity properties of exoplanets (including the SHJ model) that appeals to a
tidal realignment of the host star by a misaligned HJ.
These apparent difficulties can, however, be alleviated in the context of the
SHJ scenario and our present model.
Specifically, in the SHJ picture the realignment of the host star occurs on a
relatively long timescale ($\lesssim 1\,$Gyr; see \citealt{MatsakosKonigl15}).
This is much longer than the lifetime ($\sim$1--10\,Myr) of the gaseous disk
that gives rise to both the misaligned ``nearby'' planets and the debris disk
(which, in the scenario considered in this paper, are associated with the inner
and outer parts of the disk, respectively).
The inferred alignment properties of debris disks can be understood in this
picture if these disks are not much older than $\sim$1\,Gyr, so that the stellar
spin axis still points roughly along its original direction (which coincides
with the symmetry axis of the outer disk).
We searched the literature for age estimates of the 11 uniformly observed debris
disks tabulated in \citet{Greaves+14} and found that only two (10~CVn and
61~Vir) are definitely much older than $1$\,Gyr.
Now, \citet{MatsakosKonigl15} estimated that $\sim$50\% of systems ingest an SHJ
and should exhibit spin--orbit alignment to within $20^\circ$, with the rest
remaining misaligned.
Thus, the probability of observing an aligned debris disk in an older system is
$\sim 1/2$, implying that the chance of detecting 2 out of 2 such systems is
$\sim 1/4$.
It is, however, worth noting that the two aforementioned systems may not
actually be well aligned: based on the formal measurement uncertainties quoted
in \citet{Greaves+14}, the misalignment angle could be as large as $36^\circ$ in
10~CVn and $31^\circ$ in 61~Vir.
Further measurements that target old systems might be able to test the proposed
explanation, although one should bear in mind that additional factors may affect
the observational findings.
For example, in the tidal-downsizing scenario of planet formation, debris disks
are less likely to exist around stars that host giant planets \citep[see][]
{FletcherNayakshin16}.

\section{Conclusion}
  \label{sec:conclusion}

In this paper we conduct a proof-of-concept study of a variant of the primordial
disk misalignment model of \citet{Batygin12}.
In that model, a binary companion with an orbital radius of a few hundred au
exerts a gravitational torque on a protoplanetary disk that causes its plane to
precess and leads to a large-amplitude oscillation of the spin--orbit angle
$\theta_\mathrm{sd}$ (the angle between the angular momentum vectors of the disk
and the central star).
Motivated by recent observations, we explore an alternative model in which the
role of the distant binary is taken by a giant planet with an orbital radius of
just a few au.
Such a companion likely resided originally in the disk, and its orbit most
probably became inclined away from the disk's plane through a gravitational
interaction with other planets (involving either scattering or resonant
excitation).

Our model setup is guided by indications from numerical simulations
\citep{Xiang-GruessPapaloizou13} that, in the presence of the misaligned planet,
the disk separates at the planet's orbital radius into inner and outer parts
that exhibit distinct dynamical behaviors even as each can still be well
approximated as a rigid body.
We integrate the secular dynamical evolution equations in the quadrupole
approximation for a system consisting of the inclined planet, the two disk
parts, and the spinning star, with the disk assumed to undergo continuous mass
depletion.
We show that this model can give rise to a broad range of values for the angle
between the angular momentum vectors of the inner disk and the star (including
values of $\theta_\mathrm{sd}$ in excess of $90^\circ$), but that the
orientation of the outer disk remains virtually unchanged.
We demonstrate that the misalignment is induced by the torque that the planet
exerts on the inner disk and that it is suppressed when the mass depletion time
in the outer disk is much longer than in the inner disk, so that the outer disk
remains comparatively massive and the fast precession that it induces in the
motions of the inner disk and the planet effectively breaks the dynamical
coupling between the latter two.
Our calculations reveal that the largest misalignments are attained when the
initial disk mass is low (on the order of that of observed systems at the onset
of the transition-disk phase).
We argued that, when the misalignment angle is large, the inner and outer parts
of the disk  become fully detached and damping of the planet's orbital
inclination by dynamical friction effectively ceases.
This suggests a consistent primordial misalignment scenario: the inner region of
a protoplanetary disk can be strongly misaligned by a giant planet on a
high-inclination orbit if the disk's mass is low (i.e., late in the disk's
evolution); in turn, the planet's orbital inclination is least susceptible to
damping in a disk that undergoes a strong misalignment.

We find that, in addition to the precession-related oscillations seen in the
binary-companion model, the spin--orbit angle also exhibits a secular growth in
the planetary-companion case, corresponding to a monotonic increase in the angle
between the inner disk's and the total (inner disk plus planet) angular momentum
vectors (accompanied by a monotonic decrease in the angle between the planet's
and the total angular momentum vectors).
This behavior arises when the magnitude of the inner disk's angular momentum is
initially comparable to that of the planet but drops below it as a result of
mass depletion (on a timescale that is long in comparison with the precession
period).
This does not happen when the companion is a binary, since in that case the
companion's angular momentum far exceeds that of the inner disk at all times.
On the other hand, in the binary case the mass depletion process can drive the
system to a resonance between the disk--planet and star--disk precession
frequencies, which has the potential of significantly increasing the maximum
value of $\theta_\mathrm{sd}$ \citep[e.g.,][]{BatyginAdams13, Lai14}.
We show that this resonance is not encountered when the companion is a nearby
planet because---in contrast with the binary-companion case, in which the
disk--binary precession frequency remains constant---both of these  precession
frequencies decrease with time in the planetary-companion case. However, we
also show that when the torque that the star exerts on the disk is
taken into account (and not just that exerted by the companion, as in previous
treatments), the misalignment effect of the resonance crossing in the binary
case is measurably weaker.

A key underlying assumption of the primordial disk-misalignment model is that
the planets embedded in the disk remain confined to its plane as the disk's
orientation shifts, so that their orbits become misaligned to the same extent as
that of the gaseous disk.
However, the precession frequency that a binary companion induces in the disk
can be significantly higher than the one induced by its direct interaction with
an embedded planet, which would lead to the planet's orbital plane separating
from that of the disk: this argument was used to critique the original version
of the primordial misalignment model \citep[e.g.,][]{PicognaMarzari15}.
However, this potential difficulty is mitigated in the planetary-companion
scenario, where the ratio of these two frequencies is typically substantially
smaller.

The apparent difference in the obliquity properties of HJs around cool and hot
stars can be attributed to the tidal realignment of a cool host star by an
initially misaligned HJ \citep[e.g.,][]{Albrecht+12}.
The finding \citep{Mazeh+15} that this dichotomy is exhibited also by lower-mass
planets and extends to orbital distances where tidal interactions with the star
are very weak motivated the SHJ proposal \citep{MatsakosKonigl15}, which
postulates that $\sim$50\% of systems contain an HJ that arrives through
migration in the protoplanetary disk and becomes stranded near its inner edge
for a period of $\lesssim 1$\,Gyr---during which time the central star continues
to lose angular momentum by magnetic braking---until the tidal interaction with
the star finally causes it to be ingested (resulting in the transfer of the
planet's orbital angular momentum to the star and in the realignment of the
stellar spin in the case of cool stars).
This picture fits naturally with the primordial misalignment model discussed in
this paper.
In this broader scenario, the alignment properties of currently observed planets
(which do not include SHJs) can be explained if these planets largely remain
confined to the plane of their primordial parent disk.
In the case of cool stars the planets exhibit strong alignment on account of the
realignment action of a predecessor SHJ, whereas in the case of hot stars they
exhibit a broad range of spin--orbit angles, reflecting the primordial range of
disk misalignment angles that was preserved on account of the ineffectiveness of
the tidal realignment process in these stars.
A distinguishing prediction of the planetary-companion variant of the primordial
misalignment model in the context of this scenario arises from the expected
difference in the alignment properties of the inner and outer disks, which
implies that the good alignment exhibited by planets around cool stars should
give way to a broad range of apparent spin--orbit angles above a certain orbital
period.
There is already an observational indication of this trend \citep{LiWinn16}, but
additional data are needed to firm it up.
A complementary prediction, which is potentially also testable, is that the
range of obliquities exhibited by planets around hot stars would narrow toward 
$\lambda=0^\circ$ at large orbital periods.
This scenario may also provide an explanation for another puzzling observational
finding---that large-scale debris disks are by and large well aligned with the
spin vector of the central star---which, on the face of it, seems inconsistent
with the spin-realignment hypothesis.
In this interpretation, debris disks are associated with the outer parts of
protoplanetary disks and should therefore remain aligned with the central
star---as a general rule for hot stars, but also in the case of cool hosts that
harbor a stranded HJ if they are observed before the SHJ realigns the star.
This explanation is consistent with the fact that the great majority of observed
debris disks have inferred ages $\ll 1$\,Gyr, but the extent to which it
addresses the above finding can be tested through its prediction that a
sufficiently large sample of older systems should also contain misaligned disks.

\acknowledgements
We are grateful to Dan Fabrycky, Tsevi Mazeh, and Sean Mills for fruitful
discussions.
We also thank Gongjie Li and Josh Winn for helpful correspondence, and the
referee for useful comments.
This work was supported in part by NASA ATP grant NNX13AH56G and has made
use of NASA's Astrophysics Data System Bibliographic Services and of
\texttt{matplotlib}, an open-source plotting library for Python
\citep{Hunter07}.

\bibliographystyle{apj}
\bibliography{paper}

\appendix

\section{Calculation of the torques and precession frequencies}
  \label{app:torques}

\subsection{Torques}

\begin{figure}
  \includegraphics[width=\columnwidth]{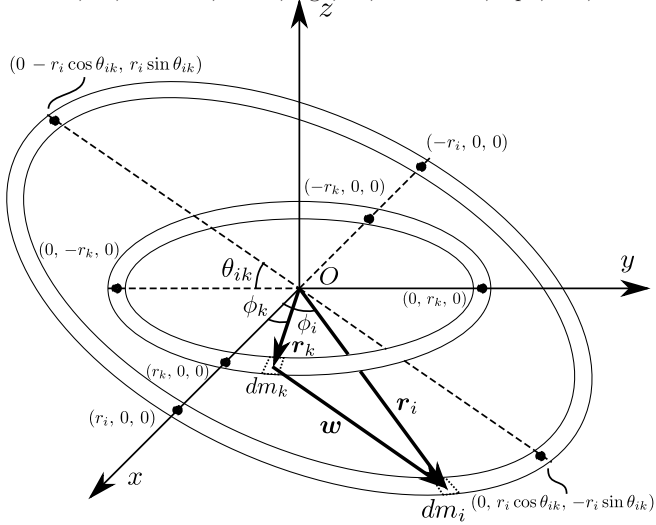}
  \caption{
    Basic configuration for the torque calculation.
    The Cartesian coordinate system is defined so that ring $k$ lies in the
    $x$--$y$ plane, with the plane containing ring $i$ intersecting it along the
    $x$ axis at an angle $\theta_{ik}$.
    The two rings are centered at $O$ and have radii $r_k$ and $r_i$,
    respectively, and mass elements $dm_k$ and $dm_i$.
    \label{fig:rings}}
\end{figure}
Figure~\ref{fig:rings} shows two concentric rings in a Cartesian coordinate
system, oriented so that their mutual gravitational torques induce a rotation
about the $x$ axis.
Because of the configuration's symmetry, the only nonzero component of the
torque that ring $i$ exerts on ring $k$ is that along the $x$ axis:
\begin{equation}
  \left[\boldsymbol{T}_{ik}\right]_x = -\int_k\int_i\left[
    \boldsymbol{y}_k\times\left(
    \frac{G\,dm_k\,dm_i}{w^2}\,\hat{\boldsymbol{w}}\right)\right]_x\,,
  \label{eq:torque}
\end{equation}
where
\begin{equation}
  \boldsymbol{y}_k = r_k\sin\phi_k\,\hat{\boldsymbol{y}}\,,
\end{equation}
\begin{equation}
  \hat{\boldsymbol{w}} = \frac{\boldsymbol{w}}{w} =
    \frac{\boldsymbol{r}_i - \boldsymbol{r}_k}{w}\,,
\end{equation}
\begin{equation}
  \boldsymbol{r}_k = r_k\cos\phi_k\,\hat{\boldsymbol{x}}
    + r_k\sin\phi_k\,\hat{\boldsymbol{y}}\,,
\end{equation}
\begin{eqnarray}
  \boldsymbol{r}_i = r_i\cos\phi_i\,\hat{\boldsymbol{x}}
    &+& r_i\sin\phi_i\cos\theta_{ik}\,\hat{\boldsymbol{y}} \nonumber\\
    &-& r_i\sin\phi_i\sin\theta_{ik}\,\hat{\boldsymbol{z}}\,,
\end{eqnarray}
\begin{eqnarray}
  w = \left[r_k^2 + r_i^2 \right.&-&\left.
    2r_kr_i\left(\cos\phi_k\cos\phi_i \right.\right. \nonumber\\
    &&\quad\quad +
    \left.\left.\sin\phi_k\sin\phi_i\cos\theta_{ik}\right)\right]^{1/2},
\end{eqnarray}
\begin{equation}
  \left[\boldsymbol{y}_k\times\boldsymbol{w}\right]_x
    = r_kr_i\sin\phi_k\sin\phi_i\sin\theta_{ik}\,\hat{\boldsymbol{x}}\,,
\end{equation}
and $\int_k$, $\int_i$ denote integrals over the masses $m_k$ and $m_i$.
These expressions can be readily generalized to a ``continuum of rings''---i.e.,
a disk---with inner and outer radii of $r_\mathrm{in}$ and $r_\mathrm{out}$,
respectively.
In the case of a ring $dm = \lambda r\,d\phi$, where $\lambda = m/2\pi r$ is the
linear mass density, whereas in the case of a disk $dm = \Sigma r\,drd\phi$,
where $\Sigma$ is the surface density.
Adopting $\Sigma = \Sigma_0(r_0/r)$ (as in \citealt{Batygin12} and
\citealt{Lai14}), where $\Sigma_0$, $r_0$ are constants, gives
$m = 2\pi\Sigma_0r_0(r_\mathrm{out} - r_\mathrm{in})$.
Therefore, $dm$ can be expressed as
\begin{equation}
  dm = \left\{\begin{array}{ll}
    \dfrac{m}{2\pi}\,d\phi & \quad\mathrm{for\ a\ ring}\,, \\ \\
    \dfrac{m}{2\pi(r_\mathrm{out} - r_\mathrm{in})}\,drd\phi &
    \quad\mathrm{for\ a\ disk}\,.
    \end{array}\right.
\end{equation}

For $r_i \gg r_k$ one can approximate
\begin{eqnarray}
  \frac{1}{w^3} \simeq \frac{1}{r_i^3}
    &+& \frac{3r_k}{r_i^4}\cos\phi_k\cos\phi_i \nonumber\\
    &+& \frac{3r_k}{r_i^4}\sin\phi_k\sin\phi_i\cos\theta_{ik}\,,
\end{eqnarray}
and thus the torque becomes
\[
  \left[T_{ik}\right]_x \simeq - A_i\,B_k\,\sin\theta_{ik}\cos\theta_{ik} =
\]
\begin{equation}
  = - \left(3G\!\!\int_i\frac{\sin^2\phi_i}{r_i^3}\,dm_i\right)\!\!
    \left(\int_kr_k^2\sin^2\phi_k\,dm_k\right)\,\sin\theta_{ik}\cos\theta_{ik}
  \label{eq:approximation}
\end{equation}
(the other terms integrate to zero), where
\begin{equation}
  A = \left\{\begin{array}{ll}
    \dfrac{3Gm}{2r^3} & \quad\quad\mathrm{for\ a\ ring}\,, \\ \\
    \dfrac{3Gm(r_\mathrm{out}+r_\mathrm{in})}
    {4r_\mathrm{out}^2r_\mathrm{in}^2} & \quad\quad\mathrm{for\ a\ disk}\,,
    \end{array}\right.
\end{equation}
and
\begin{equation}
  B = \left\{\begin{array}{ll}
    \dfrac{mr^2}{2} & \quad\quad\mathrm{for\ a\ ring}\,, \\ \\
    \dfrac{m(r_\mathrm{out}^3-r_\mathrm{in}^3)}
    {6(r_\mathrm{out} - r_\mathrm{in})} & \quad\quad\mathrm{for\ a\ disk}\,.
    \end{array}\right.
\end{equation}
The torque that $k$ exerts on $i$ is $[T_{ki}]_x = -[T_{ik}]_x$.\footnote{
Note that the value of $[T_{ki}]_x$ cannot be calculated from
Equation~(\ref{eq:approximation}), which only holds for $r_i \gg r_k$, and
instead has to be evaluated from $[T_{ik}]_x$ using Newton's third law.}
Equation~(\ref{eq:approximation}) can also be used when object $k$ is a star by
setting $B_k = k_qM_*R_*^2\Omega_*^2/(GM_*/R_*^3)$ and, in the case of a
protostar, using $k_q \simeq 0.1$ (the value appropriate to fully convective
stars; e.g., \citealt{Lai14}).

\begin{figure*}
  \begin{center}
  \begin{tabular}{ccc}
    \includegraphics[width=0.65\columnwidth]{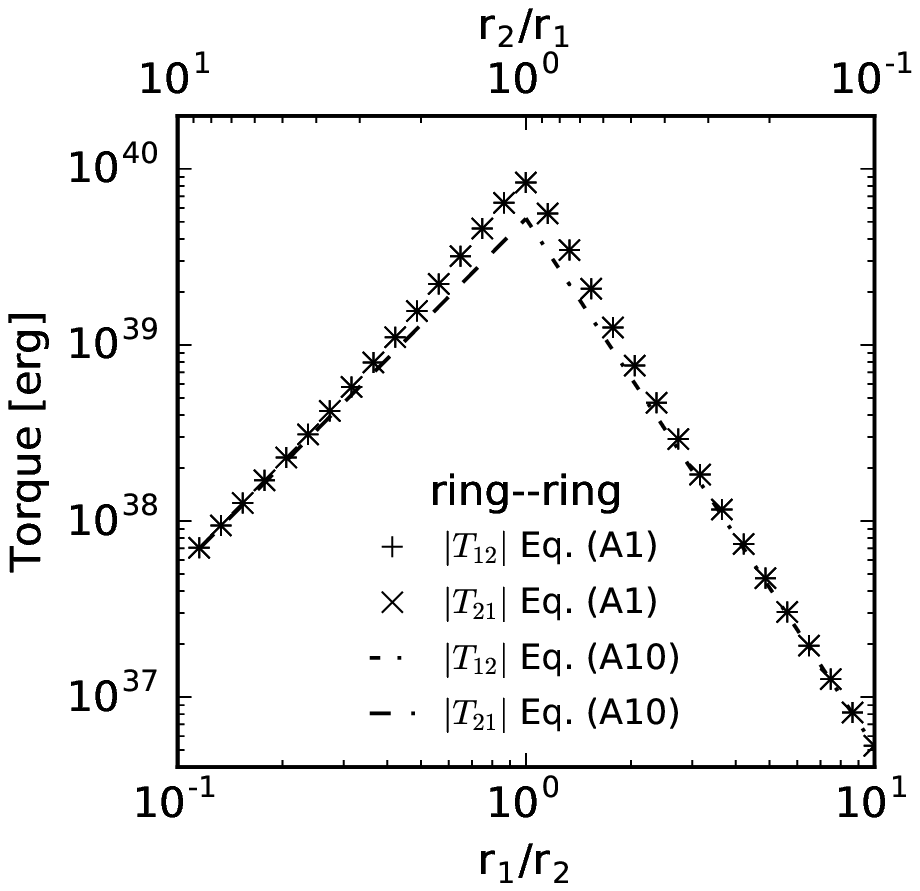} &
    \includegraphics[width=0.65\columnwidth]{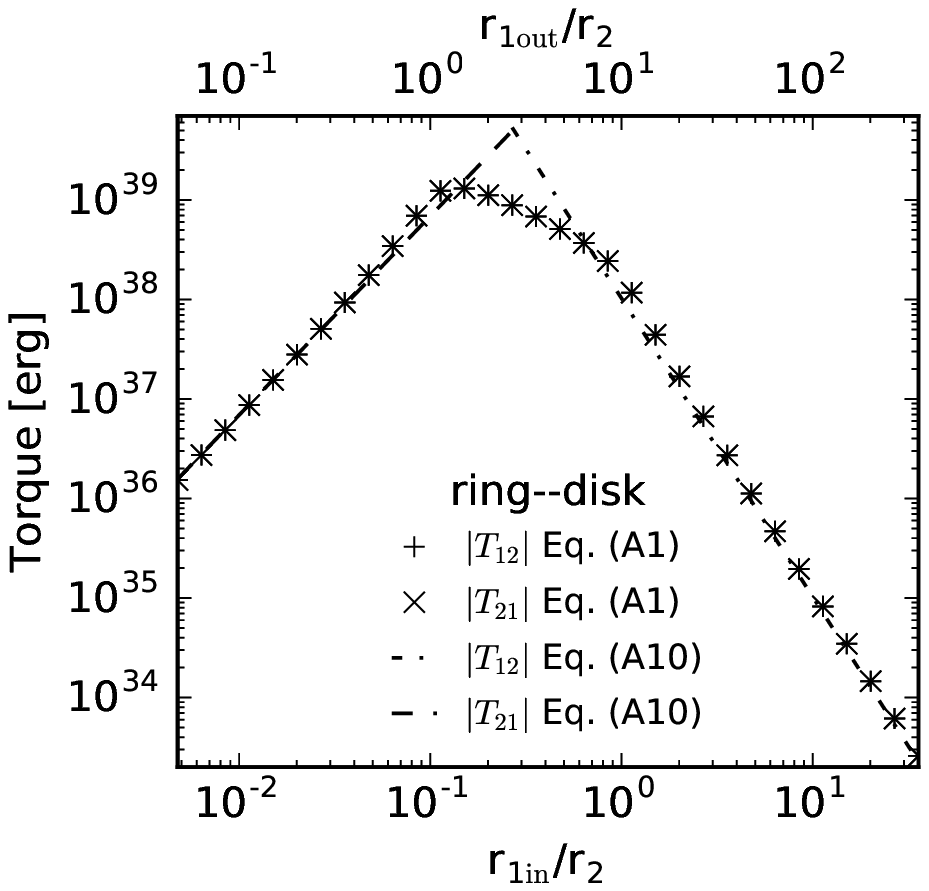} &
    \includegraphics[width=0.65\columnwidth]{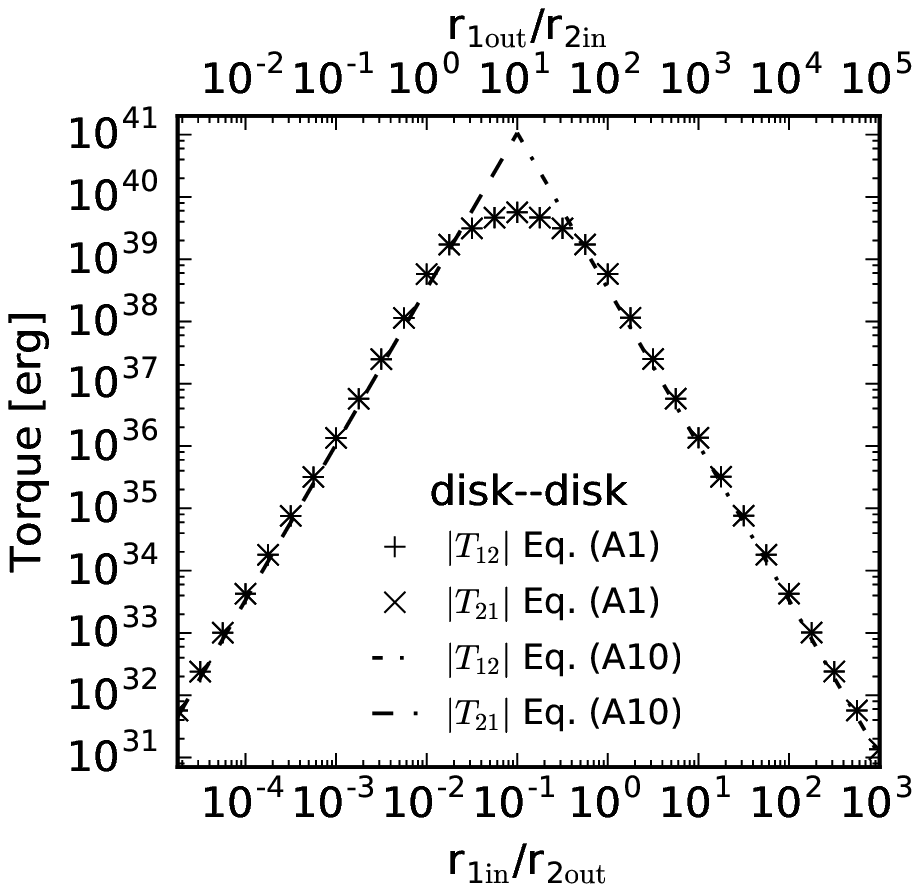}
  \end{tabular}
  \end{center}
  \caption{
    Comparison of the exact torque (Equation~(\ref{eq:torque}); points) with the
    quadrupole approximation (Equation~(\ref{eq:approximation}); lines) for
    three generic configurations: ring--ring (left-hand panel), ring--disk
    (middle panel), and disk--disk (right-hand panel).
    Each pair of symbols ($+$ and $\times$, corresponding to $T_{12}$ and
    $T_{21}$, respectively) represents a different system, with the ratio(s) of
    their defining radii shown on the top and bottom horizontal axes.
  \label{fig:approximation}}
\end{figure*}
Figure~\ref{fig:approximation} compares the torque that is calculated using
Equation~(\ref{eq:torque}) (points) with the approximate expression of
Equation~(\ref{eq:approximation}) (lines) for the following systems: two rings
(left-hand panel), a ring and a disk (middle panel), and two disks (right-hand
panel).
In all cases the mass of each object is taken to be $m = M_\mathrm{J}$, and we
set $\theta_{ik} = 30^\circ$.
Each point in the figure (representing a superposed pair of $+$ and $\times$
symbols) corresponds to a different system, characterized by its relevant
parameters (the radius of the ring or the inner and outer radii of the disk).
For the ring--ring system, ring 1 has radius $r_1 = 1$\,au, and the different
cases correspond to $r_2\in[0.1,\,10]$\,au.
For the ring--disk system, the ring has radius $r_2 = 2.8$\,au, and the inner
and outer edges of the disk lie in the ranges
$r_{1\mathrm{in}}\in[0.013,\,133]$\,au and
$r_{1\mathrm{out}}\in[0.13,\,1,333]$\,au, respectively.
The same ranges are adopted for both disks in the case of the disk--disk system.
The figure indicates that for the cases that are relevant to the present
study---in particular, when a ring (representing a planet) is located at either
the inner or the outer edge of the disk, or when the two disks are
adjacent---Equation~(\ref{eq:approximation}) provides a very good approximation
to the torque.

Our approach, which amounts to using only the lowest-order (quadrupole) term in
the expansion of the interaction potential \citep[e.g.,][]{PapaloizouTerquem95},
is less general than the Gaussian averaging method employed by \cite{Batygin12},
but, as demonstrated in Figure~\ref{fig:approximation}, it is entirely adequate
for our purposes.
In our treatment of the disk we make the further approximations that its
constituent ``rings'' remain circular and that it behaves as a rigid body.
In the case of a binary companion and $M_\mathrm{t} = 0.01\,M_*$,
\citet{Batygin12} verified by an explicit calculation that these approximations
are well justified even if the binary moves on an eccentric orbit and
self-gravity is the only mode of internal interaction in the disk.

\subsection{Precession Frequencies}

By combining Equations~(\ref{eq:precession}) and~(\ref{eq:approximation}) we
obtain an analytic expression for the precession frequencies:
\begin{equation}
  \Omega_{ik} = \Omega_{ki} \simeq
  - A_i\,B_k\,\frac{J_{ik}}{L_iL_k}\cos\theta_{ik}\,,
\end{equation}
where, again, object $i$ is taken to be ``outside of'' object $k$
($r_i \gg r_k$).
The six characteristic frequencies are
\begin{eqnarray}
\label{eq:omega_sd}
  \Omega_\mathrm{sd} &\simeq& -4.86\times10^{-5}\,
    \left(\frac{2k_q}{k_*}\right)
    \left(\frac{M_\mathrm{d}}{0.01M_*}\right)\nonumber\\
    &\times&\left(\frac{R_*}{2R_\odot}\right)
    \left(\frac{r_\mathrm{d,in}}{4R_*}\right)^{-2}
    \left(\frac{r_\mathrm{d,out}}{5\,\mathrm{au}}\right)^{-1}
    \frac{J_\mathrm{sd}}{D}\nonumber\\
    &\times&\left(\frac{\Omega_*}{0.1\sqrt{GM_\odot/(2R_\odot)^3}}\right)
    \cos\theta_\mathrm{sd}\,\frac{2\pi}{\mathrm{yr}}\,,
\end{eqnarray}
\begin{eqnarray}
\label{eq:omega_sp}
  \Omega_\mathrm{sp} &\simeq& -5.14\times10^{-10}\,
    \left(\frac{2k_q}{k_*}\right)
    \left(\frac{M_\mathrm{p}}{M_\mathrm{J}}\right)
    \left(\frac{M_*}{M_\odot}\right)^{-1}\nonumber\\
    &\times&\left(\frac{R_*}{2R_\odot}\right)^3
    \left(\frac{a}{5\,\mathrm{au}}\right)^{-3}
    \frac{J_\mathrm{sp}}{P}\nonumber\\
    &\times&\left(\frac{\Omega_*}{0.1\sqrt{GM_\odot/(2R_\odot)^3}}\right)
    \cos\theta_\mathrm{sp}\,\frac{2\pi}{\mathrm{yr}}\,,
\end{eqnarray}
\begin{eqnarray}
\label{eq:omega_sh}
  \Omega_\mathrm{sh} &\simeq& -2.42\times10^{-9}\,
    \left(\frac{2k_q}{k_*}\right)
    \left(\frac{M_\mathrm{h}}{0.09M_*}\right)\nonumber\\
    &\times&\left(\frac{R_*}{2R_\odot}\right)^3
    \left(\frac{r_\mathrm{h,in}}{5\,\mathrm{au}}\right)^{-2}
    \left(\frac{r_\mathrm{h,out}}{50\,\mathrm{au}}\right)^{-1}
    \frac{J_\mathrm{sh}}{H} \nonumber\\
    &\times&\left(\frac{\Omega_*}{0.1\sqrt{GM_\odot/(2R_\odot)^3}}\right)
    \cos\theta_\mathrm{sh}\,\frac{2\pi}{\mathrm{yr}}\,,
\end{eqnarray}
\begin{eqnarray}
\label{eq:omega_dp}
  \Omega_\mathrm{dp} &\simeq& -2.23\times10^{-4}\,
    \left(\frac{M_\mathrm{d}}{0.01M_*}\right)\nonumber\\
    &\times&\left(\frac{r_\mathrm{d,out}}{a}\right)^2
    \frac{J_\mathrm{dp}}{D}\nonumber\\
    &\times&\left(\frac{\Omega_\mathrm{p}}
    {\sqrt{GM_\odot/(5\,\mathrm{au})^3}}\right)
    \cos\theta_\mathrm{dp}\,\frac{2\pi}{\mathrm{yr}} \\
    &\simeq& -3.20\times10^{-5}\,
    \left(\frac{M_\mathrm{p}}{M_\mathrm{J}}\right)
    \left(\frac{M_*}{M_\odot}\right)^{-1}\nonumber\\
    &\times&\left(\frac{r_\mathrm{d,out}}{a}\right)^{3/2}
    \frac{J_\mathrm{dp}}{P}\nonumber\\
    &\times&\left(\frac{\Omega_\mathrm{p}}
    {\sqrt{GM_\odot/(5\,\mathrm{au})^3}}\right)
    \cos\theta_\mathrm{dp}\,\frac{2\pi}{\mathrm{yr}}\,,
\end{eqnarray}
\begin{eqnarray}
\label{eq:omega_dh}
  \Omega_\mathrm{dh} &\simeq& -1.51\times10^{-3}\,
    \left(\frac{M_\mathrm{h}}{0.09M_*}\right)\nonumber\\
    &\times&\left(\frac{r_\mathrm{d,out}}{r_\mathrm{h,in}}\right)^2
    \left(\frac{r_\mathrm{d,out}}{r_\mathrm{h,out}}\right)
    \frac{J_\mathrm{dh}}{H}\nonumber\\
    &\times&\left(\frac{GM_*/r_\mathrm{d,out}^3}
    {GM_\odot/(5\,\mathrm{au})^3}\right)^{1/2}
    \cos\theta_\mathrm{dh}\,\frac{2\pi}{\mathrm{yr}}\,,
\end{eqnarray}
and
\begin{eqnarray}
\label{eq:omega_ph}
  \Omega_\mathrm{ph} &\simeq& -3.02\times10^{-3}\,
    \left(\frac{M_\mathrm{h}}{0.09M_*}\right)\nonumber\\
    &\times&\left(\frac{a}{r_\mathrm{h,in}}\right)^2
    \left(\frac{a}{r_\mathrm{h,out}}\right)
    \frac{J_\mathrm{ph}}{H}\nonumber\\
    &\times&\left(\frac{GM_*/a^3}{GM_\odot/(5\,\mathrm{au})^3}\right)^{1/2}
    \cos\theta_\mathrm{ph}\,\frac{2\pi}{\mathrm{yr}}\,.
\end{eqnarray}

\section{Angular momentum of the inner disk and planet}
  \label{app:Jdp}

To obtain an expression for the time evolution of
$\boldsymbol{J}_\mathrm{dp} = \boldsymbol{D} + \boldsymbol{P}$, we write
\begin{equation}
  \frac{d\boldsymbol{D}}{dt} = \boldsymbol{T}_\mathrm{pd}
    + \left(\frac{d\boldsymbol{D}}{dt}\right)_\mathrm{depl}\,,
\end{equation}
\begin{equation}
  \frac{d\boldsymbol{P}}{dt} = -\boldsymbol{T}_\mathrm{pd}\,,
\end{equation}
and take their sum using Equation~(\ref{eq:dDdt}):
\begin{eqnarray}
  \frac{d\boldsymbol{J}_\mathrm{dp}}{dt}
    = \left(\frac{dD}{dt}\right)_\mathrm{depl}(\cos\phi'\sin\theta'\hat{x}'
    &+& \sin\phi'\sin\theta'\hat{y}' \nonumber\\
    &+& \cos\theta'\hat{z}')\,,
\end{eqnarray}
where we expressed $\hat{\boldsymbol{D}}$ in a cartesian coordinate system
$(x',\,y',\,z')$ with $\hat{z}' = \hat{\boldsymbol{J}}_\mathrm{dp}$ and with
$\theta'$, $\phi'$ the spherical polar angles.
Since the precession period is much shorter than the depletion time (for
example, for the parameters that characterize model~\texttt{DP-M}, the initial
value of $\tau_\mathrm{dp}/\tau$ is $\simeq 0.017$), it is an excellent
approximation to treat $(dD/dt)_\mathrm{depl}$ as a constant over one precession
period.
Averaging over $\phi$ therefore gives
\begin{equation}
  \left<\frac{d\boldsymbol{J}_\mathrm{dp}}{dt}\right> \simeq
    \left(\frac{dD}{dt}\right)_\mathrm{depl}\hat{\boldsymbol{J}}_\mathrm{dp}\,,
\end{equation}
where the angle brackets denote an average over a precession period.
This implies that $\left<\boldsymbol{J}_\mathrm{dp}\right>$ decreases in
magnitude without changing its direction.
The oscillation of $\psi_\mathrm{j}$ during a single precession
period---described in Figure~\ref{fig:vectors}---is in practice so small (its
amplitude is $\simeq \tau_\mathrm{dp}/\tau$ for $t \ll \tau$) that it cannot be
picked out in Figures~\ref{fig:DP-M} and~\ref{fig:DP-m}.

\section{Resonance Crossing in Star--Disk--Binary Systems}
  \label{app:resonance}

\begin{figure}
  \includegraphics[width=\columnwidth]{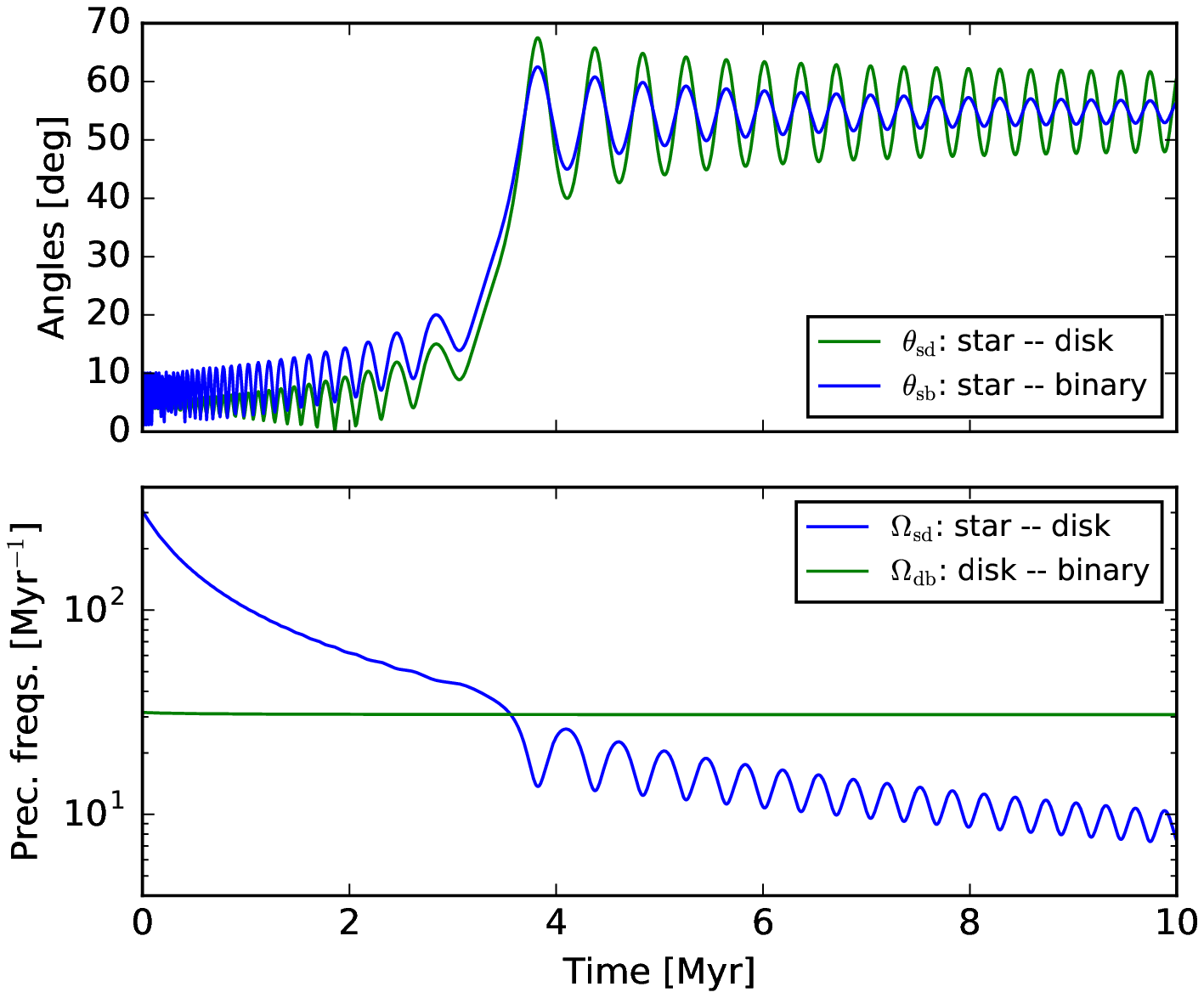}
  \includegraphics[width=\columnwidth]{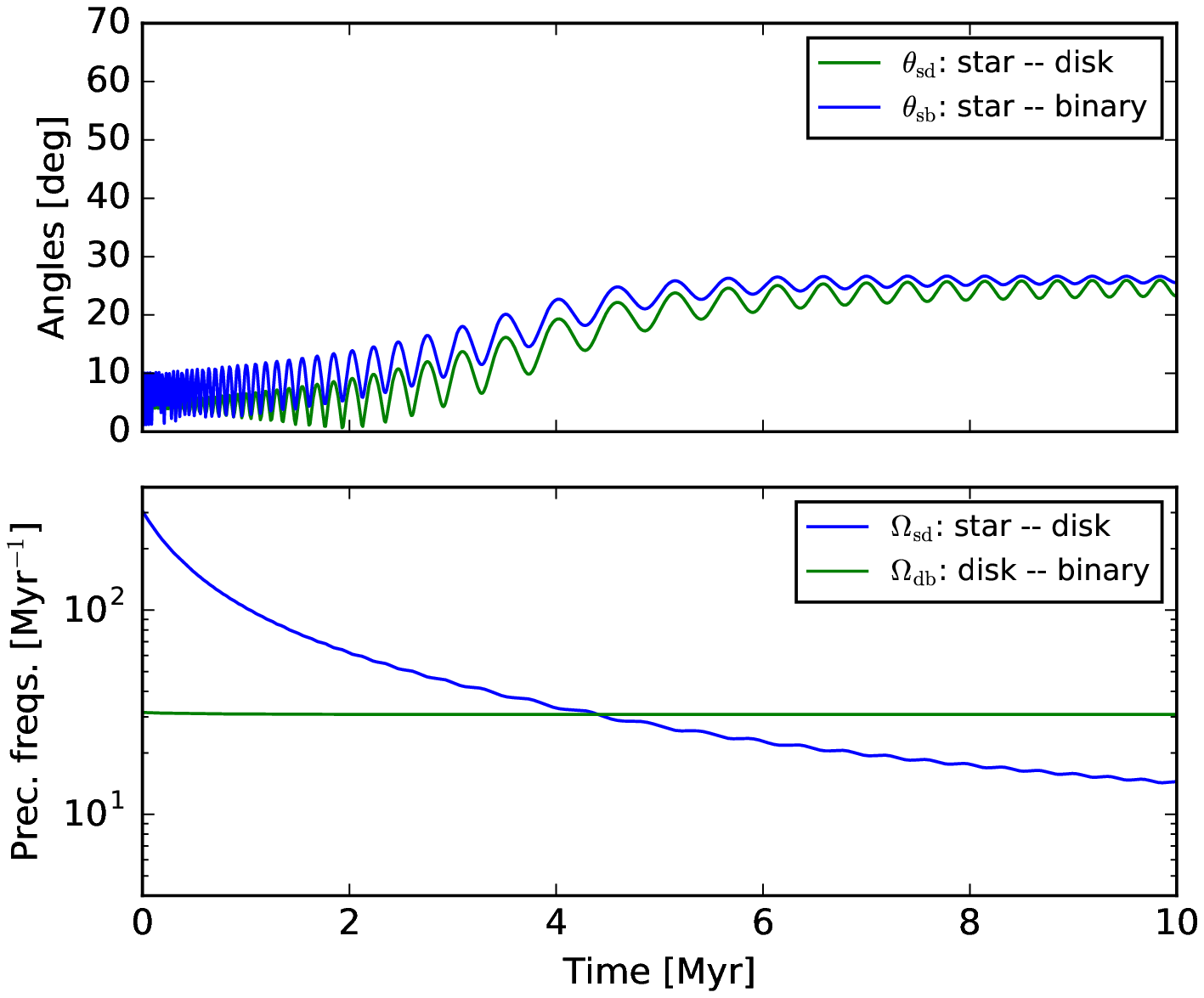}
  \caption{
    Time evolution of a star--disk--binary system (model \texttt{binary}) for
    the parameters used in figure~3 of \citet{Lai14}.
    The top two panels present the results obtained by neglecting the torque
    that the star exerts on the disk, whereas in the bottom two panels the
    effect of this torque is included.
  \label{fig:binary}}
\end{figure}
The formalism employed in this paper can also be used to treat the original
variant of the primordial disk misalignment model, in which the companion is a
distant star rather than a nearby giant planet.
To validate our code, we consider one such system (model \texttt{binary}; see
Table~\ref{tab:models}), which corresponds to the example presented in figure~3
of \citet{Lai14}.
In that work, the evolution of a system consisting of a star, a binary star
(subscript b), and a disk that undergoes mass depletion (according to the
prescription given by Equation~(\ref{eq:deplete})) was studied by integrating
the equations
\begin{equation}
  \label{eq:Lai1}
  \frac{d\boldsymbol S}{dt} = \boldsymbol T_\mathrm{ds}
\end{equation}
and 
\begin{equation}
  \label{eq:Lai2}
  \frac{d\boldsymbol D}{dt} = \boldsymbol T_\mathrm{bd}\,.
\end{equation}
The results of solving these two equation with our numerical scheme are
presented in the top two panels of Figure~\ref{fig:binary}.
These results are identical to those obtained by \citet{Lai14} and indicate that
even a system with small initial misalignments between the star and the disk
($\theta_\mathrm{sd} = 5^\circ$) and between the disk and the binary
($\theta_\mathrm{db} = 5^\circ$, with $\theta_\mathrm{sb} = 10^\circ$) can
attain a large final spin--orbit angle if a resonance between the precessions
frequencies $\Omega_\mathrm{sd}$ and $\Omega_\mathrm{db}$ is crossed.
In this case the precession frequency that the torque exerted by the disk
induces in the stellar angular momentum vector is initially high enough
($\Omega_\mathrm{sd} > \Omega_\mathrm{db}$) for the star--disk pair to remain
coupled as the disk precesses under the influence of the binary.
However, as the mass of the disk becomes depleted, $\Omega_\mathrm{sd}$
decreases and eventually crosses $\Omega_\mathrm{db}$.
Beyond that point, the stellar angular momentum can no longer follow the
precession of the disk's angular momentum, and the motion of these two vectors
decouples.
At resonance the star--disk system may attain a large misalignment, which, in
the absence of strong star--disk coupling, remains ``frozen'' during the ensuing
evolution.

In the model presented in \citet{Lai14} the torque that the star exerts on the
disk is neglected.
We now use the more general formulation employed in this work to extend that
model by including also the torques exerted by the star.
Thus, instead of Equations~(\ref{eq:Lai1}) and~(\ref{eq:Lai2}), we integrate
\begin{equation}
  \frac{d\boldsymbol S}{dt} =
  \boldsymbol T_\mathrm{ds} + \boldsymbol T_\mathrm{bs}
\end{equation}
and
\begin{equation}
  \frac{d\boldsymbol D}{dt} =
  \boldsymbol T_\mathrm{sd} + \boldsymbol T_\mathrm{bd}\,.
\end{equation}
In practice, only the torque that the star exerts on the disk plays a role, with
the effect of $\boldsymbol T_\mathrm{bs}$ remaining negligible (see
Equations~(\ref{eq:omega_sd}) and~(\ref{eq:omega_sp})).
The results of this integration are shown in the bottom two panels of
Figure~\ref{fig:binary} and demonstrate that the back torque that the star
exerts on the disk can significantly reduce the effectiveness of the resonance
misalignment mechanism, so that its effect cannot in general be neglected.

\end{document}